\documentclass[12pt]{article}

\usepackage{epsfig}

\usepackage{graphicx}

\def\square{\kern1pt\vbox{\hrule height 1.2pt
\hbox{\vrule width 1.2pt\hskip 3pt
\vbox{\vskip 6pt}\hskip 3pt\vrule width 0.6pt}
\hrule height 0.6pt}\kern1pt}
\def\ltwid{\mathrel{\raise.3ex\hbox{$<$\kern-.75em\lower1ex\hbox{$\sim$}}}}

\begin{document}

\begin{titlepage}
\begin{flushright}
UFIFT-QG-16-07, CCTP-2016-09 \\
CCQCN-2016-146, ITCP-IPP 2016/07
\end{flushright}

\vspace{0.5cm}

\begin{center}
\bf{An Improved Cosmological Model}
\end{center}

\vspace{0.3cm}

\begin{center}
N. C. Tsamis$^{\dagger}$
\end{center}
\begin{center}
\it{Institute of Theoretical \& Computational Physics, \\
Department of Physics, University of Crete \\
GR-710 03 Heraklion, HELLAS.}
\end{center}

\vspace{0.2cm}

\begin{center}
R. P. Woodard$^{\ast}$
\end{center}
\begin{center}
\it{Department of Physics, University of Florida \\
Gainesville, FL 32611, UNITED STATES.}
\end{center}

\vspace{0.3cm}

\begin{center}
ABSTRACT
\end{center}
\hspace{0.3cm} We study a class of non-local, action-based,
and purely gravitational models. These models seek to describe
a cosmology in which inflation is driven by a large, bare
cosmological constant that is screened by the self-gravitation
between the soft gravitons that inflation rips from the vacuum.
Inflation ends with the universe poised on the verge of
gravitational collapse, in an oscillating phase of expansion
and contraction that should lead to rapid reheating when matter
is included. After the attainment of a hot, dense universe the
nonlocal screening terms become constant as the universe evolves
through a conventional phase of radiation domination. The onset
of matter domination triggers a much smaller anti-screening
effect that could explain the current phase of acceleration.

\vspace{0.3cm}

\begin{flushleft}
\end{flushleft}

\vspace{0.1cm}

\begin{flushleft}
$^{\dagger}$ e-mail: tsamis@physics.uoc.gr \\
$^{\ast}$ e-mail: woodard@phys.ufl.edu
\end{flushleft}

\end{titlepage}

\section{Introduction}

On scales larger than about $100 Mpc$ the universe is 
well described by the geometry:
\begin{equation}
ds^2 \; = \;
- dt^2 \, + \, a^2(t) \, d{\bf x} \cdot d{\bf x}
\;\; . \label{frw}
\end{equation}
The time variation of the scale factor $a(t)$ gives 
the instantaneous values of the Hubble parameter $H(t)$  
and the deceleration parameter $q(t)$ or, equivalently, 
the first slow-roll parameter $\epsilon$:
\begin{eqnarray}
H(t) &\!\! \equiv \!\!&
\frac{\dot{a}(t)}{a(t)}
\; = \; \frac{d}{dt} \ln a(t)
\;\; , \label{H} \\
q(t) &\!\! \equiv \!\!&
- \frac{\dot{a}(t) \; \ddot{a}(t)}{\dot{a}^2(t)}
\; = \;
-1 - \frac{\dot{H}(t)}{H^2(t)}
\; \equiv \;
-1 + \epsilon(t)
\;\; . \label{q}
\end{eqnarray}
Their current values are:
$H_0 \simeq (67.8 \pm 0.9) km/sec \, Mpc $ and 
$\epsilon_0 \simeq 0.462 \pm 0.017$ \cite{Planck2015}.
\footnote{In quoting these numbers we have used fits 
from cosmic ray microwave data which effectively
exploits the $\Lambda$CDM model for $z \sim 1000$.
Larger and significantly different values for $H_0$
arise from Hubble plots which exploit $\Lambda$CDM
for $z \sim 1$ \cite{riess}.}

There is overwhelming evidence that the history 
of the universe included a period of very early 
($t \sim 10^{-33}sec$) accelerated expansion known 
as inflation and defined by $H > 0$ with $\epsilon 
< 1$ \cite{Linde,Slava}. During the inflationary era 
infrared gravitons are produced out of the vacuum 
because of the accelerated expansion of spacetime 
\cite{gravitons}. The interaction stress among the 
gravitons produced -- an inherently non-local effect 
-- can lead to a non-trivial quantum gravitational 
back-reaction on inflation \cite{NctRpw1}. General 
counting rules give the following leading infrared
behaviour for the Hubble parameter $H(t)$ at late 
times in de Sitter spacetime \cite{NctRpw2}:
\begin{equation}
H(t) \, = \,
H_{\rm in} \, \Big\{ 1 - G \Lambda \Big( 
c_2 G \Lambda \ln[a(t)]
+ c_3 (G \Lambda)^2 \ln^2[a(t)] 
+ \dots \Big) \Big\} 
\;\; . \label{Hpert}
\end{equation}
It becomes evident from (\ref{Hpert}) that 
perturbation theory breaks down when $\ln[a(t)] 
\sim (G \Lambda)^{-1}$ and that evolving beyond 
this point requires non-perturbative techniques.

In the absence of non-perturbative results, it is 
perhaps desirable to propose phenomenological models 
that can provide calculable evolution beyond 
perturbation theory \cite{nonloc}. This can be 
accomplished by modifying the field equations:
\begin{equation}
G_{\mu\nu} + \Delta G_{\mu\nu} [g] 
\; = \;
- \Lambda \, g_{\mu\nu}
\;\; , \label{eom}
\end{equation}
where $\Delta G_{\mu\nu} [g]$ encodes the full effect
of the quantum-induced gravitational back-reaction.
\footnote{Hellenic indices take on spacetime values 
while Latin indices take on space values. 
Our metric tensor $g_{\mu\nu}$ has spacelike signature 
and our curvature tensor equals:
$R^{\alpha}_{~\beta\mu\nu} \equiv 
\Gamma^{\alpha}_{~\nu\beta, \mu} +
\Gamma^{\alpha}_{~\mu\rho} \;
\Gamma^{\rho}_{~\nu\beta} -
(\mu \leftrightarrow \nu)$. 
The initial Hubble constant is $3H^2_{\rm in} \equiv 
\Lambda$. We restrict our analysis to scales 
$M \equiv (\, \Lambda / 8 \pi G \,)^{\frac14}$ 
below the Planck mass $M_{\rm Pl} \equiv G^{-\frac12}$ 
so that the dimensionless coupling constant 
$G \Lambda$ of the theory is small.}

Any such model should: \\
${\bullet \;}$ Be consistent with the perturbative 
expectation (\ref{Hpert}). \\
${\bullet \;}$ Reflect the non-local nature of the 
back-reaction effect in a causal way. \\
${\bullet \;}$ Respect stress-energy conservation. \\
${\bullet \;}$ Not disturb the basic ability of the 
gravitational equations to evolve from the initial 
spacelike surface with knowledge of the metric and 
its first time derivative only. \\
The hope is that the actual construction of the model 
will contain the most cosmologically significant part 
of the full effective quantum gravitational equations.

Previously \cite{NctRpw3, NctRpw4} we proposed a 
phenomenological model based on an effective conserved 
stress-energy tensor $T_{\mu\nu}[g]$:
\begin{equation}
\Delta G_{\mu\nu} [g] \, = \,
- 8 \pi G \, T_{\mu\nu}[g] 
\;\; , \label{eom1}
\end{equation}
which takes the perfect fluid form:
\begin{equation}
T_{\mu\nu}[g] \, = \,
\Big( \rho[g] + p[g] \Big) u_{\mu}[g] \, u_{\nu}[g]
+ p[g] \, g_{\mu\nu}
\quad , \quad
g^{\mu\nu} u_{\mu}[g] \, u_{\nu}[g] = -1
\;\; , \label{Tmn}
\end{equation}
where the ansatz for the pressure $p[g]$ is:
\begin{equation}
p[g] \, = \,
\Lambda^2 \, f(Y)
\quad , \quad
Y \, \equiv \, 
- G \Lambda \, \frac{1}{\square} R
\;\; . \label{p}
\end{equation}
It can be shown \cite{NctRpw3} that all models of this 
generic type where the function $f(Y)$ grows monotonically 
and without bound: \\
${\bullet \;}$ Experience a long phase of inflation. \\
${\bullet \;}$ The end of inflation leads to a short 
phase of oscillations. \\
${\bullet \;}$ The participation of all super-horizon 
modes to the oscillations furnishes a natural and very 
fast reheating mechanism for the cosmos using only the 
universal gravitational coupling to matter. \\
${\bullet \;}$ If matter couplings that allow energy
dissipation are added it is plausible that the epoch 
of oscillations ends in a radiation domination epoch. \\ 
However, these models have negative attributes as well: \\
${\bullet \;}$ There is a ``sign problem'' because their 
post-inflationary evolution eventually makes the pressure
positive and thus in conflict with the observed late time
acceleration \cite{riess2, wang}. \\
${\bullet \;}$ There is a ``magnitude problem'' because 
the magnitude of the total pressure produced by the source 
is unacceptably large relative to the current pressure. 

There is another generic class of models where the source
of $\Delta G_{\mu\nu}[g]$ is a quantum-induced non-local
effective action term $\Delta S[g]$:
\begin{equation}
\Delta G_{\mu\nu}[g] \, = \,
\frac{16 \pi G}{\sqrt{-g}} \;
\frac{\delta \Delta S[g]}{\delta g^{\mu\nu}}
\quad , \quad
\Delta S \, \equiv \, 
\int \! d^4x \,\, \Delta {\cal L}[g]
\;\; , \label{DS}
\end{equation}
where we parametrize $\Delta {\cal L}[g]$ as follows:
\begin{equation}
\Delta {\cal L}[g] \; = \;
\Lambda^2 \, h \! \left( X[g] \right) \sqrt{-g}
\;\; . \label{DL}
\end{equation}

The purpose of this paper is to present a phenomenological
model of the latter kind that does not suffer from the
sign and magnitude problems described above. Section 2 
describes the construction of the model and derives the
dynamical equations it satisfies. Section 3 presents 
numerical and semi-analytical results for the cosmology 
predicted. Our conclusions comprise Section 4.

\section{The Model}

Simple tools to construct a reasonable {\it ansatz} 
for the most cosmologically significant part of the 
full effective action are: \\
${\bullet \;}$ Curvature invariants whose specialization 
to the geometry (\ref{frw}) gives:
\begin{eqnarray}
& \mbox{} &
R \, = \, 6 ( 2 - \epsilon ) H^2
\quad , \quad 
R^2 \, = \, 36 ( 2 - \epsilon )^2 H^4
\;\; , \label{R} \\
& \mbox{} &
R_{\mu\nu} R^{\mu\nu} \, = \,
12 ( 3 - 3\epsilon + \epsilon^2 ) H^4 
\;\; , \label{Rmn} \\
& \mbox{} &
R_{\mu\nu\rho\sigma} R^{\mu\nu\rho\sigma} \, = \,
12 ( 2 - 2\epsilon + \epsilon^2 ) H^4 
\;\; , \label{Rmnrs}
\end{eqnarray}
${\bullet \;}$ Invariant differential operators whose 
inverses can plausibly introduce non-locality and whose
specialization to the geometry (\ref{frw}) gives:
\begin{eqnarray}
\square &\!\! = \!\!&
\frac{1}{\sqrt{-g}} \, \partial_{\mu} \left( 
\sqrt{-g} \, g^{\mu\nu} \, \partial_{\nu} \right)
\, = \,
-\partial_t^2 - 3H \partial_t
\;\; , \label{square} \\
\square_c &\!\! = \!\!&
\square - \frac16 \, R
\, = \,
-\partial_t^2 - 3H \partial_t - 2H^2 - {\dot H}
\;\; , \label{squarec}
\end{eqnarray}
when acting on a function of co-moving time.
Their inverses:
\begin{eqnarray}
\frac{1}{\square} &\!\! = \!\!&
\int_{t_{\rm in}}^t dt' \frac{1}{a^3(t')} 
\int_{t_{\rm in}}^{t'} dt'' \, a^3(t'')
\;\; , \label{invsquare} \\
\frac{1}{\square_c} &\!\! = \!\!&
\frac{1}{a(t)} \int_{t_{\rm in}}^t dt' \frac{1}{a(t')} 
\int_{t_{\rm in}}^{t'} dt'' \, a^2(t'')
\;\; , \label{invsquarec}
\end{eqnarray}
are defined with retarded boundary conditions to avoid 
the appearance of new degrees of freedom \cite{deser}.

A way of achieving the desired properties for the
induced source $X[g]$ in (\ref{DL}) is as follows: \\
${\bullet \;}$ To address the ``magnitude problem'',
we must move the very high scale factor of $\Lambda 
= 3H_{\rm in}^2$ which appears in (\ref{p}) to the 
right and have it re-appear essentially as $H^2(t)$, 
a factor that decreases in magnitude like $t^{-2}$
after inflation:
\begin{equation}
Y \, = \, 
- 3 G H_{\rm in}^2 \, \frac{1}{\square} R
\quad \rightarrow \quad
- 3 G \, \frac{1}{\square} H^2 R
\;\; . \label{magnitude}
\end{equation}
An immediate consequence is that our ansatz now 
requires an extra curvature invariant -- see 
(\ref{R}-\ref{Rmnrs}) -- to account for the extra 
factor of $H^2$. \\
${\bullet \;}$ To address the ``sign problem'', 
$X[g]$ must change sign as we exit the inflationary 
epoch ($\epsilon < 1$) and enter the post-inflationary 
epoch ($\epsilon > 1$). By inspecting expressions 
(\ref{R}-\ref{Rmnrs}) we conclude that the combination:
\begin{equation}
\frac13 R^2 - R_{\mu\nu} R^{\mu\nu} 
\, = \,
12 ( 1 - \epsilon ) H^4 
\;\; , \label{sign}
\end{equation}
indeed changes sign as $\epsilon$ passes through 1. \\
${\bullet \;}$ The requirement for the effect to 
become quiescent during radiation domination 
($\epsilon =2$) is most easily satisfied by having 
the curvature scalar $R$ present in the ansatz. \\
${\bullet \;}$ The ansatz must contain an {\it overall}
factor of $\square^{-1}$ to account for the secular
nature of the effect which implies the need for an
operator with memory for the behaviour of the source
by not extinguishing its effect as the universe 
evolves. Furthermore, $\square^{-1}$ provides the 
single infrared logarithm dictated by the de Sitter 
correspondence limit (\ref{Hpert}) to order 
$(G\Lambda)^2$. \\
${\bullet \;}$ The dimensionality of the ansatz 
for $X[g]$ requires the presence of a second inverse 
differential operator. To preserve the correspondence 
limit (\ref{Hpert}) this operator must not give an
additional infrared logarithm to order $(G\Lambda)^2$
and it is $\square_c^{-1}$ that has this property. 

Therefore, the proposed {\it ansatz} consists of 
the following quantum-induced source $X[g]$:
\footnote{An alternate choice would have been: 
$X[g] = G \square^{-1} \square_c^{-1} R \left( 
\frac13 R^2 - R_{\mu\nu} R^{\mu\nu} \right)$.
It is equally well-motivated and may have interesting
comological evolution.}
\begin{equation}
X[g] \; \equiv \;
G \, \frac{1}{\square} \, R \, \frac{1}{\square_c}
\left( \frac13 \, R^2 - R_{\mu\nu} R^{\mu\nu} \right)
\;\; . \label{X}
\end{equation}
The function $h(X[g])$ must have the ability to end
inflation which implies that it must have the ability
to become singular.
\footnote{In the models defined by (\ref{p}) the end
of inflation was achieved by the source monotonically
increasing without bound leading, unfortunately, to 
the magnitude problem.}
The contribution $\Delta G_{\mu\nu}$ to the gravitational
field equations (\ref{eom}) is quite complicated and is 
most easily derived by going to the equivalent scalar 
representation \cite{odintsov, NctRpw5}. We introduce 
two auxiliary scalar fields $A$ and $C$ which we require 
to obey the following equations of motion:
\begin{eqnarray}
\square_c A &\!\! = \!\!&
\frac13 R^2 - R_{\mu\nu} R^{\mu\nu} 
\;\; , \label{eomA} \\
\square C &\!\! = \!\!& 
R \, A
\;\; . \label{eomC}
\end{eqnarray}
This can be achieved by introducing two Lagrange multipliers
$B$ and $D$ this way:
\begin{eqnarray}
\Delta {\cal L} &\!\! = \!\!&
\Lambda^2 \, h(G C) \sqrt{-g} \, + \,
B \left[ \square_c A - \Big( \frac13 R^2 - 
R_{\mu\nu} R^{\mu\nu} \Big) \right] \sqrt{-g} 
\nonumber \\
& \mbox{} &
+ \, D \Big[ \square C - R \, A \Big] \sqrt{-g}
\;\; , \label{DLabcd}
\end{eqnarray}
so that the desired equations of motion (\ref{eomA}-\ref{eomC})
emerge:
\begin{eqnarray}
\frac{1}{\sqrt{-g}} 
\frac{\delta (\Delta S)}{\delta B}
&\!\! = \!\!&
\square_c A - \Big( \frac13 R^2 
- R_{\mu\nu} R^{\mu\nu} \Big) \, = \, 0 
\;\; , \label{eomA2} \\
\frac{1}{\sqrt{-g}} 
\frac{\delta (\Delta S)}{\delta D}
&\!\! = \!\!&
\square C - R \, A \, = \, 0
\;\; , \label{eomC2} 
\end{eqnarray}
as well as those for the Lagrange multipliers:
\begin{eqnarray}
\frac{1}{\sqrt{-g}} 
\frac{\delta (\Delta S)}{\delta A}
&\!\! = \!\!&
\square_c B - R \, D = 0
\quad \Rightarrow \quad
B = \frac{1}{\square_c} R \, D
\;\; , \label{eomB} \\
\frac{1}{\sqrt{-g}} 
\frac{\delta (\Delta S)}{\delta C}
&\!\! = \!\!&
\square D + G \Lambda^2 h'(GC) = 0
\quad \Rightarrow \quad
D = - \frac{1}{\square} G \Lambda^2 h'(GC) 
\;\; . \qquad \label{eomD} 
\end{eqnarray}

The resulting quantum induced stress tensor should
be covariantly conserved:
\begin{equation}
T_{\mu\nu}[g] \, \equiv \,
\frac{2}{\sqrt{-g}}
\frac{\delta (\Delta S)}{\delta g^{\mu\nu}}
\qquad \Rightarrow \qquad
D^{\mu} T_{\mu\nu} = 0
\;\; , \label{DTmn}
\end{equation}
and a tedious but straightforward computation confirms 
this.

For the spacetimes of cosmological interest (\ref{frw}),
besides (\ref{R}-\ref{Rmnrs}), we have:
\begin{eqnarray}
& \mbox{} &
R_{00} \, = \, -3 ( H^2 + 6 {\dot H} )
\quad , \quad
R_{ij} \, = \, ( 3H^2 + {\dot H} ) g_{ij}
\;\; , \label{R00} \\
& \mbox{} &
D_0 \, D_0 \, = \, \partial_t^2
\quad , \quad
D_i D_j \, = \, - g_{ij} H \partial_t
\;\; . \label{D00}
\end{eqnarray}
Using all these relations, the specialization to 
(\ref{frw}) of the equations of motion for the 
auxiliary scalar fields (\ref{eomA}-\ref{eomC}) and 
the Lagrange multipliers (\ref{eomB}-\ref{eomD}) take 
the form:
\begin{eqnarray}
{\ddot A} &\!\! = \!\!&
- 3H{\dot A} - (2 - \epsilon) H^2 A
- 12 (1 - \epsilon) H^4
\;\; , \label{ddA} \\
{\ddot B} &\!\! = \!\!&
- 3H{\dot B} - (2 - \epsilon) H^2 B
- 6 (2 - \epsilon) H^2 D
\;\; , \label{ddB}\\
{\ddot C} &\!\! = \!\!&
- 3H{\dot C} - 6 (2 - \epsilon) H^2 A
\;\; , \label{ddC} \\
{\ddot D} &\!\! = \!\!&
- 3H{\dot D} + G \Lambda^2 h'(GC)
\;\; . \label{ddD}
\end{eqnarray}
The full cosmological equations (\ref{eom}) become 
for the $(00)$ component:
\begin{eqnarray}
\frac{3H^2}{16 \pi G} + \frac12 \Lambda^2 h(GC)
&\!\!\! - \!\!\!&
\frac12 \Big( {\dot A}{\dot B} + {\dot C}{\dot D}
\Big) - 6 H^3 {\dot B} 
\nonumber \\
&\!\!\! - \!\!\!& 
3( H \partial_t + H^2 ) \Big( \frac16 AB + AD \Big)
\; = \, \frac{\Lambda}{16 \pi G}
\;\; , \label{eom00}
\end{eqnarray}
and for the $(ij)$ component:
\begin{eqnarray}
& - \!\!\!\!&
(3 - 2\epsilon) \frac{H^2}{16 \pi G}
- \frac12 \Lambda^2 h(GC) + G \Lambda^2 A h'(GC)
- \frac16 {\dot A}{\dot B} - \frac12 {\dot C}{\dot D}
+ 2 {\dot A}{\dot D}
\nonumber \qquad \\
& \mbox{} &
- 2 (1 + 2\epsilon) H^3 {\dot B}
- 2 (3 - 2\epsilon) Η^4 \Big( B + 6D \Big)
- (H \partial_t + H^2) \Big( \frac16 AB + AD \Big)
\nonumber \\
& \mbox{} &
= \, 
- \frac{\Lambda}{16 \pi G}
\;\; , \label{eomij}
\end{eqnarray}
where we have used (\ref{ddA}-\ref{ddD}) to reach 
(\ref{eomij}). Moreover, it is useful to record the 
sum $(00)+(ij)$ of these two equations:
\footnote{For instance, one can again check and 
verify stress-energy conservation: \\
$\partial_t [ (00) ] = - 3H [ (00)+(ij) ]$.}
\begin{eqnarray}
\frac{2\epsilon H^2}{16 \pi G} 
&\!\!\! + \!\!\!&
G \Lambda^2 A h'(GC) - \frac23{\dot A}{\dot B}
- {\dot C}{\dot D} + 2 {\dot A}{\dot D} 
- 4 ( 2 - \epsilon ) H^3 {\dot B} 
\nonumber \\
& \mbox{} &
\hspace{-1cm}
- 2 ( 3 - 2\epsilon ) H^4 \Big( B + 6D \Big)
- 4 ( H \partial_t + H^2 ) \Big( \frac16 AB + AD \Big)
= 0
\;\; . \qquad \label{eom00+ij}
\end{eqnarray}

To describe the evolution of the model it would have 
been very convenient to use the number of e-foldings 
$n$ as the time evolution parameter. However, evolution 
with increasing $n$ is good only as long as the universe
expands; if at some point the universe stops expanding
or contracts, $n$ cannot describe the evolution because
the scale factor $a(n) = \exp(n)$ always increases with
increasing $n$.

We shall therefore use the time variable $t$ as our
evolution parameter at the cost of a more complicated
system of dynamical equations.
\footnote{Also, to economize on writing we drop the 
$16 \pi$ in front of $G$. Note that there should have 
been a $16 \pi$ in the various factors of $G$ that 
appear in the quantum source $X[g]$ (\ref{X}).}
Because these equations will be analyzed numerically 
it makes sense to use dimensionless variables. If the 
initial value of the Hubble parameter is $H(t_{\rm in}) 
\equiv H_{\rm in}$, we first define the dimensionless 
time $\tau$:
\begin{equation}
\tau \, \equiv \, H_{\rm in} \, t
\quad \Rightarrow \quad 
\partial_t \, = \, H_{\rm in} \, \partial_{\tau}
\qquad , \qquad 
' \, \equiv \, \frac{d}{d\tau}
\;\; , \label{tau}
\end{equation}
and then the remaining dimensionless variables:
\begin{eqnarray}
H^2 &\!\!\! \equiv \!\!\!& \frac{\chi^2}{G}
\quad \Rightarrow \quad
H_{\rm in}^2 \, \equiv \, \frac{\chi_{\rm in}^2}{G}
\quad , \quad
\Lambda \, \equiv \, 3 H_{\rm in}^2 \, \equiv \, 
\frac{3 \chi_{\rm in}^2}{G}
\;\; , \label{chi} \\
A &\!\!\! \equiv \!\!\!& - \frac{3 \alpha}{G}
\quad , \quad
B \, \equiv \, -3 \beta
\quad , \quad
C \, \equiv \, \frac{9 \gamma}{G}
\quad , \quad
D \, \equiv \, \delta
\;\; , \label{abcd} \\
h(GC) &\!\!\! = \!\!\!& h(9\gamma) \, \equiv \, f(\gamma)
\;\; , \label{f} \\
h'(GC) &\!\!\! \equiv \!\!\!&
\frac{\partial}{\partial(GC)} h(GC)
\, = \, 
\frac19 \, \frac{\partial}{\partial\gamma} f(\gamma)
\, \equiv \, \frac19 f'(\gamma)
\;\; . \label{f'}
\end{eqnarray}
The set of dimensionless variables we wish to solve for
is $\{ \alpha, \beta, \gamma, \delta, \chi, \epsilon \}$ 
and the initial conditions at $\tau = \tau_{\rm in}$ are:
\footnote{The initial value data (\ref{ivdabcd}) -- which
follow because $\{ \alpha, \beta, \gamma, \delta \}$ all 
equal to expressions with overall $\square^{-1}$ or 
$\square_c^{-1}$ in front -- ensure that no additional 
degrees of freedom are introduced by these four fields.}
\begin{eqnarray}
& \mbox{} &
\alpha = \alpha' \, = \, 
\beta = \beta' \, = \,
\gamma = \gamma' \, = \,
\delta = \delta' \, = \, 0
\;\; , \label{ivdabcd} \\
& \mbox{} &
\chi \, = \, \chi_{\rm in}
\quad , \quad
\epsilon \, = \, 0
\;\; . \label{ivdxe}
\end{eqnarray}
The time evolution of $\{ \alpha, \beta, \gamma, \delta \}$
is obtained from the dimensionless form of equations
(\ref{ddA}-\ref{ddD}):
\begin{eqnarray}
& \mbox{} &
\alpha'' + 3 \frac{\chi}{\chi_{\rm in}} \, \alpha'
+ ( 2 - \epsilon ) \frac{\chi^2}{\chi_{\rm in}^2} \, \alpha
\, = \,
4 ( 1 - \epsilon ) \frac{\chi^4}{\chi_{\rm in}^2}
\;\; , \label{dda} \\
& \mbox{} &
\beta'' + 3 \frac{\chi}{\chi_{\rm in}} \, \beta'
+ ( 2 - \epsilon ) \frac{\chi^2}{\chi_{\rm in}^2} \, \beta
\, = \,
2 ( 2 - \epsilon ) \frac{\chi^2}{\chi_{\rm in}^2} \, \delta
\;\; , \label{ddb}\\
& \mbox{} &
\gamma'' + 3 \frac{\chi}{\chi_{\rm in}} \, \gamma'
\, = \,
2 ( 2 - \epsilon ) \frac{\chi^2}{\chi_{\rm in}^2} \, \alpha
\;\; , \label{ddc} \\
& \mbox{} &
{\ddot \delta} + 3 \frac{\chi}{\chi_{\rm in}} \, {\dot \delta}
\, = \,
\chi_{\rm in}^2 \, f'(\gamma)
\;\; . \label{ddd}
\end{eqnarray}
Furthermore, the variable $\chi$ is solved from
the dimensionless form of the $(00)$ equation 
(\ref{eom00}):
\begin{eqnarray}
\Big[ 2 \chi_{\rm in} \, \beta' \Big] \, \chi^3
&\!\! + \!\!&
\Bigg[ \frac13 - \frac12 \alpha \beta + \alpha \delta 
\Bigg] \, \chi^2 
\, + \,
\chi_{\rm in} \, \partial_{\tau} 
\Big[ -\frac12 \alpha \beta + \alpha \delta \Big] \, \chi
\nonumber \\
&\!\! - \!\!&
\chi_{\rm in}^2 \Bigg[ \frac13 
- \frac12 \chi_{\rm in}^2 \, f(\gamma)
+ \frac12 ( \alpha' \beta' + \gamma' \delta' ) \Bigg]
\, = \, 0
\;\; . \label{eomchi}
\end{eqnarray}
This is a cubic algebraic equation which always has 
a real solution. 
\footnote{Had we been able to use $n$ as the evolution 
parameter, the resulting equation analogous to 
(\ref{eomchi}) would have been quadratic in $\chi^2$.}
The real solution of (\ref{eomchi}) which is consistent
with the correspondence limit of small values for the
coefficients of $\chi^3$ and $\chi$ is rather complicated:
\begin{equation}
\chi \, = \,
\frac{1}{3M} \,
\Big\{ -1 + \sqrt{1 - 3 M N} \,
2 \cos \! \Big[ \frac{\pi}{3} - \frac13 \arctan Q 
\Big] \Big\}
\;\; , \label{chisolution}
\end{equation}
where we have defined:
\begin{eqnarray}
M &\!\! = \!\!&
\frac{2 \chi_{\rm in} \, \beta'}
{\frac13 - \frac{\alpha \beta}{2} + \alpha \delta}
\;\; , \label{M} \\
N &\!\! = \!\!&
\frac{\chi_{\rm in} \Big( 
- \frac{\alpha' \beta}{2} 
- \frac{\alpha \beta'}{2}
+ \alpha' \delta + \alpha \delta' \Big)}
{\frac13 - \frac{\alpha \beta}{2} + \alpha \delta}
\;\; , \label{N} \\
P &\!\! = \!\!&
- \frac{\chi_{\rm in}^2 \Big[ 
\frac13 - \frac12 \chi_{\rm in}^2 \, f(\gamma)
+ \frac12 (\alpha' \beta' + \gamma' \delta') \Big]}
{\frac13 - \frac{\alpha \beta}{2} + \alpha \delta}
\;\; , \label{P} \\
Q &\!\! = \!\!&
\frac{3 \sqrt{3} M 
\sqrt{-P + \frac{N^2}{4} + \frac92 M N P
- M N^3 - \frac{27}{4} M^2 P^2}}
{1 - \frac92 M N + \frac{27}{2} M^2 P}
\;\; . \label{Q}
\end{eqnarray}
Moreover, we solve for $\epsilon$ from the dimesionless
form of the $(00)+(ij)$ equation (\ref{eom00+ij}):
\begin{eqnarray}
& \mbox{} &
\hspace{-1.5cm}
\Big[ 2 \chi^2 + 12 \chi_{\rm in} \, \chi^3 \beta'
+ 12 \chi^4 ( -\beta + 2\delta ) \Big] \, \epsilon
\, = \, 
3 \, \Big\{ \chi_{\rm in}^4 \, \alpha f'(\gamma)
\nonumber \\
& \mbox{} &
+ \chi_{\rm in}^2 ( \, 2\alpha' \beta' + 3\gamma' \delta' 
+ 2\alpha' \delta' \, )
- 8\chi_{\rm in} \, \chi^3 \beta' 
- 6 \chi^4 ( \beta - 2\delta )
\nonumber \\
& \mbox{} &
+ 4 ( \chi_{\rm in} \, \chi \, \partial_{\tau} + \chi^2 )
\Big( \frac12 \alpha \beta - \alpha \delta \Big) \Big\}
\;\; . \label{eomepsilon}
\end{eqnarray}
Finally, it will be useful for the analysis to follow
to combine the equations (\ref{eomchi}, \ref{eomepsilon})
for $\chi$ and $\epsilon$ and make the simplifications 
that arise: 
\begin{eqnarray}
\epsilon \chi^2 &\!\! = \!\!&
\frac{3}{2 + \beta' \chi_{\rm in} \, \chi 
+ 12 (-\beta + 2 \delta) \chi^2} 
\times 
\label{chieps} \\
& \mbox{} &
\hspace{-1cm}
\Big\{ 
[ \alpha f'(\gamma) + 2 f(\gamma) ] \, \chi_{\rm in}^4
- \frac43 \chi_{\rm in}^2
+ ( 2\alpha' + \gamma' ) \, \delta' \chi_{\rm in}^2
+ \frac43 \chi^2 + 6 ( -\beta + 2 \delta ) \, \chi^4
\Big\}
\;\; . \nonumber
\end{eqnarray}

Finally, we must make a choice for the function $f(X)$. 
The perturbative result (\ref{Hpert}) indicates that 
the effect gets strong when $G \Lambda \, H_{\rm in} 
\, t \sim 1$, and this corresponds to $X \sim 1$. 
A simple appropriate singular algebraic function is:
\begin{equation}
f(X[g]) \, = \, 
\frac{X[g]}{1 - X[g]} \, = \,
\frac{1}{1 - X[g]} - 1
\;\; , \label{fchoice}
\end{equation}
which also has the property that the small $X$ limit 
of $f(X)$ is $X$.

\section{The Resulting Cosmology}

The purpose of this section is to describe the sort 
of background cosmology this model produces. We begin 
with a discussion of inflation and how it ends, then 
we describe the immediate post-inflationary phase. These 
portions of the treatment are supported by substantial 
numerical analysis, reported in the form of graphs. 
Subsequent evolution involves matter in an essential 
way, so we limit the discussion to some general comments. 

\subsection{The Inflationary Regime}

\begin{figure}[ht]
\includegraphics[width=6cm,height=4.8cm]{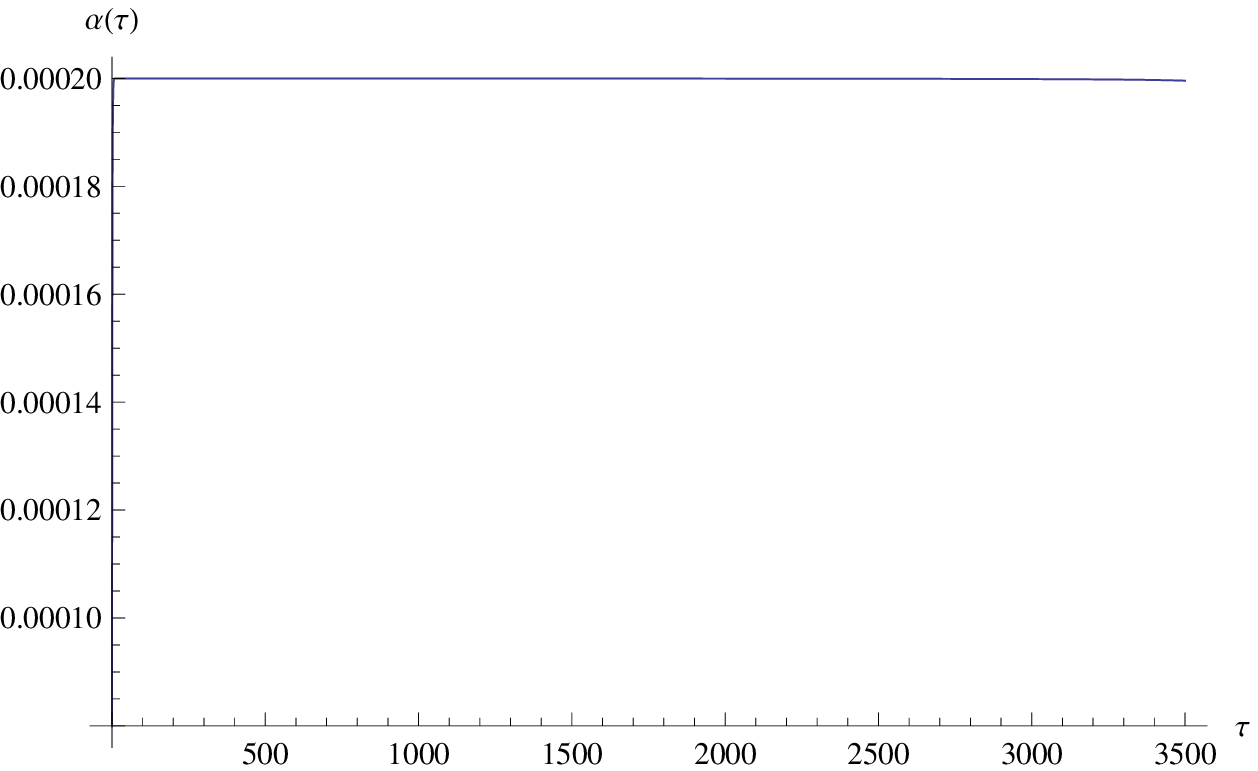}
\hspace{1cm}
\includegraphics[width=6cm,height=4.8cm]{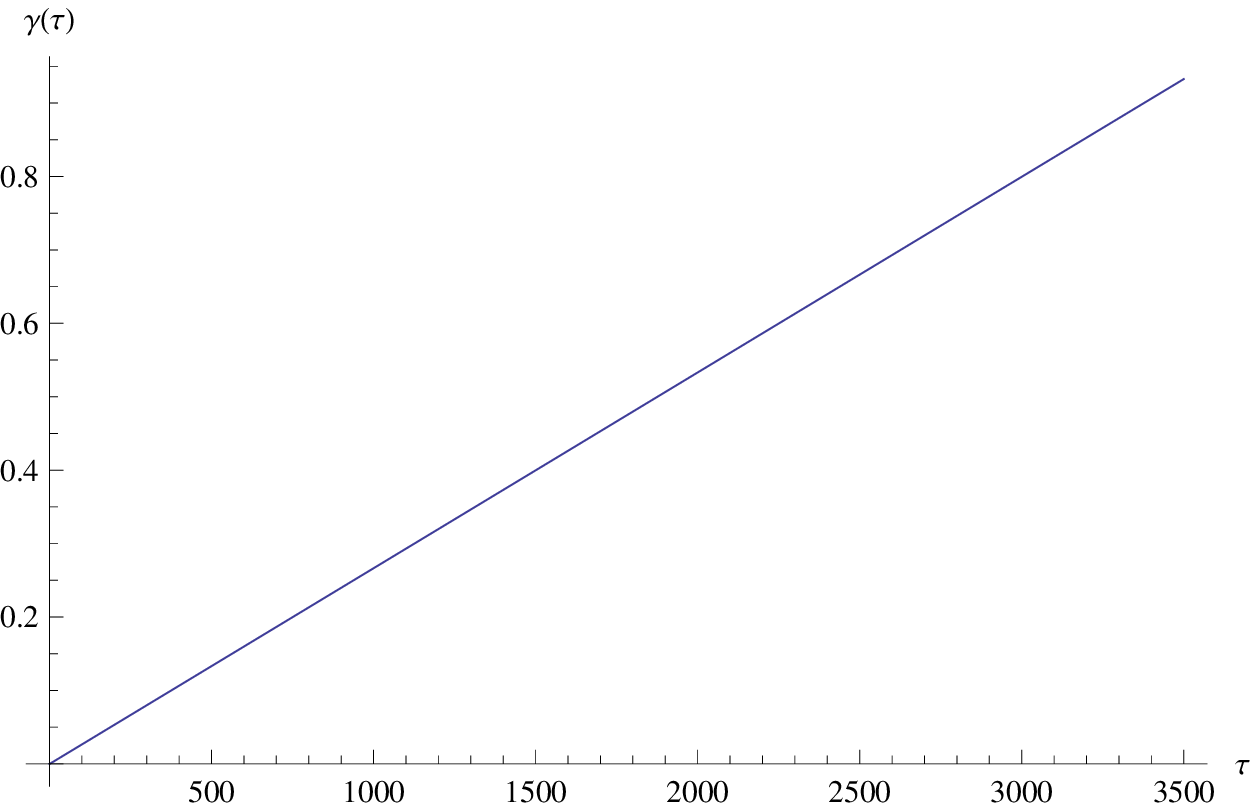}
\caption{\footnotesize Numerical simulation of the auxiliary 
scalar functions $\alpha(\tau)$ and $\gamma(\tau)$ for 
$\chi_{\rm in} = \frac1{100}$ and $f(X) = \frac{X}{1-X}$. Note 
the quantitative agreement with expressions (\ref{earlyalpha}) 
and (\ref{earlygamma}) during the full epoch of de Sitter 
expansion.}
\label{alphagamma3500}
\end{figure}

\begin{figure}[ht]
\includegraphics[width=6cm,height=4.8cm]{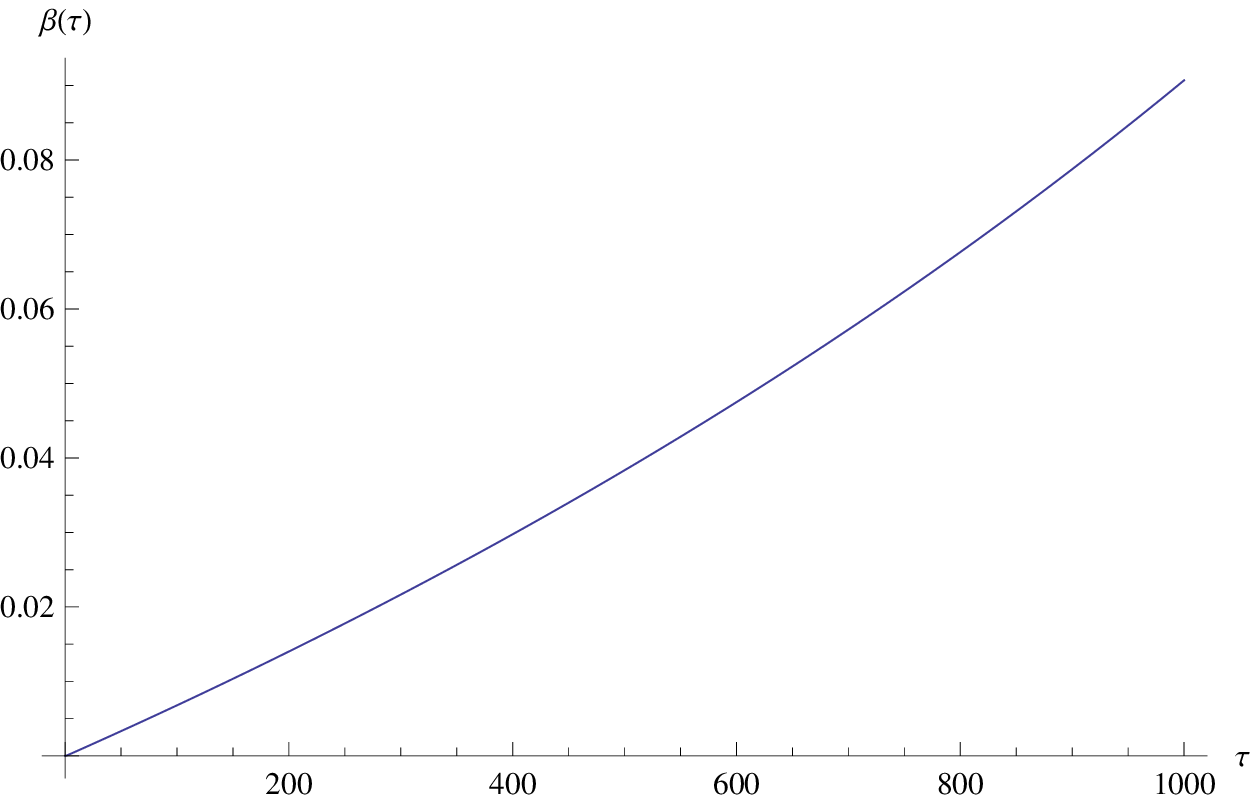}
\hspace{1cm}
\includegraphics[width=6cm,height=4.8cm]{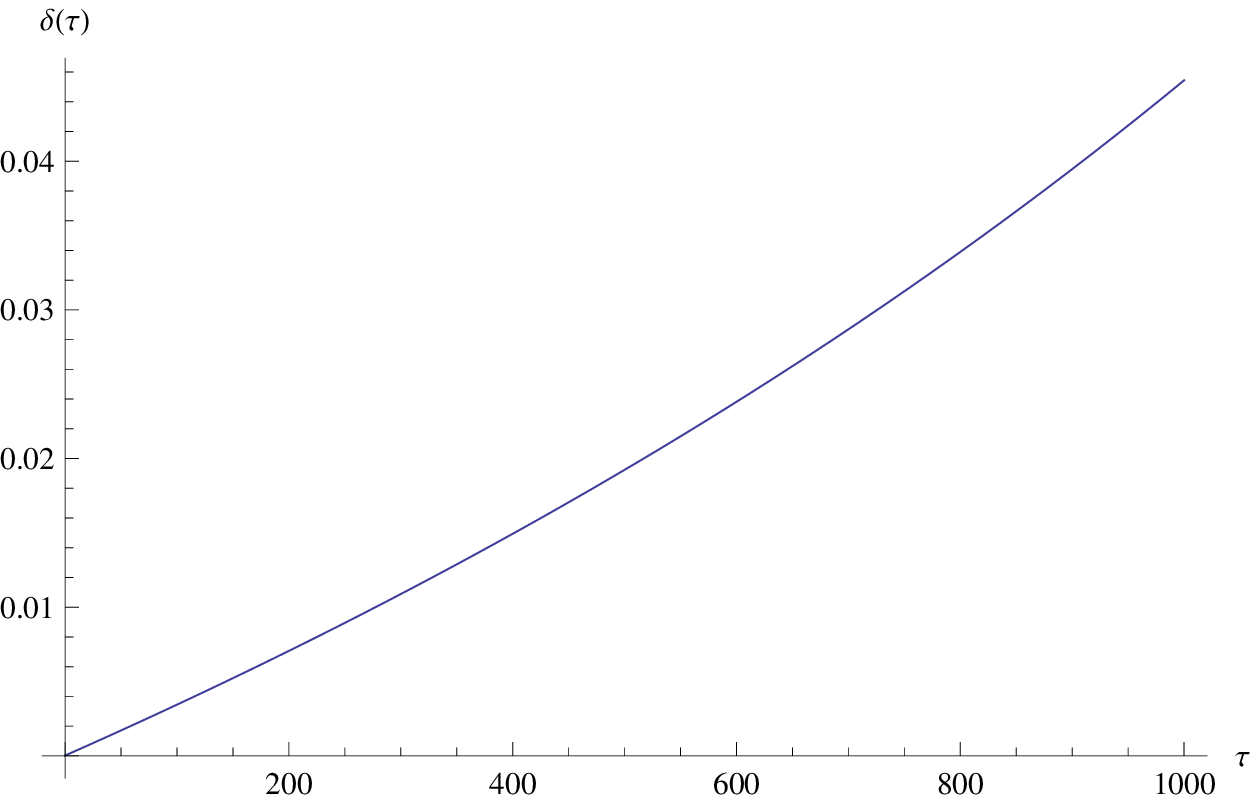}
\caption{\footnotesize Numerical simulation of the auxiliary 
scalar functions $\beta(\tau)$ and $\delta(\tau)$ for 
$\chi_{\rm in} = \frac1{100}$ and $f(X) = \frac{X}{1-X}$. Note 
the rough agreement with expressions (\ref{earlybeta}) and 
(\ref{earlydelta}) during the early epoch of de Sitter expansion.}
\label{betadelta1000}
\end{figure}

For a long period the scale factor remains at nearly 
its de Sitter value of $a(\tau) = e^{\tau}$. During 
this phase the four scalars experience some minor 
transients which decay like powers of $e^{-\tau}$ 
to reveal forms which persist until screening becomes 
significant:
\begin{eqnarray}
\alpha(\tau) &\!\! = \!\!& 
2 \chi_{\rm in}^2 \Bigl( 1 \!-\! e^{-\tau}\Bigr)^2 
\; \longrightarrow \; 
2 \chi_{\rm in}^2 
\; , \label{earlyalpha} \\
\beta(\tau) &\!\! = \!\!& 
\frac23 \chi_{\rm in}^2 \Bigl( \tau \!-\! \frac{11}{6}
\!+\! 3 e^{-\tau} \!-\! \frac32 e^{-2\tau} 
\!+\! \frac13 e^{-3\tau} \Bigr) 
\; \longrightarrow \;
\frac23 \chi_{\rm in}^2 \Bigl( \tau \!-\! \frac{11}{6} \Bigr) 
\; , \label{earlybeta} \\
\gamma(\tau) &\!\! = \!\!& 
\frac83 \chi_{\rm in}^2 \Bigl( \tau \!-\! \frac{11}{6}
\!+\! 3 e^{-\tau} \!-\! \frac32 e^{-2\tau} 
\!+\! \frac13 e^{-3\tau} \Bigr) 
\; \longrightarrow \;
\frac83 \chi_{\rm in}^2 \Bigl( \tau \!-\! \frac{11}{6}
\Bigr) \; , \label{earlygamma} \\
\delta(\tau) &\!\! = \!\!& 
\frac13 \chi_{\rm in}^2 \Bigl( \tau \!-\! \frac{1}{3}
\!+\! \frac{1}{3} e^{-3\tau} \Bigr) 
\; \longrightarrow \; 
\frac13 \chi_{\rm in}^2 
\Bigl( \tau \!-\! \frac{1}{3} \Bigr) 
\; . \label{earlydelta}
\end{eqnarray}
These behaviours are evident in Figures~\ref{alphagamma3500} 
and \ref{betadelta1000}, which were generated for 
$\chi_{\rm in} = \frac1{100}$ and $f(X) = \frac{X}{1-X}$.

\begin{figure}[ht]
\includegraphics[width=6cm,height=4.8cm]{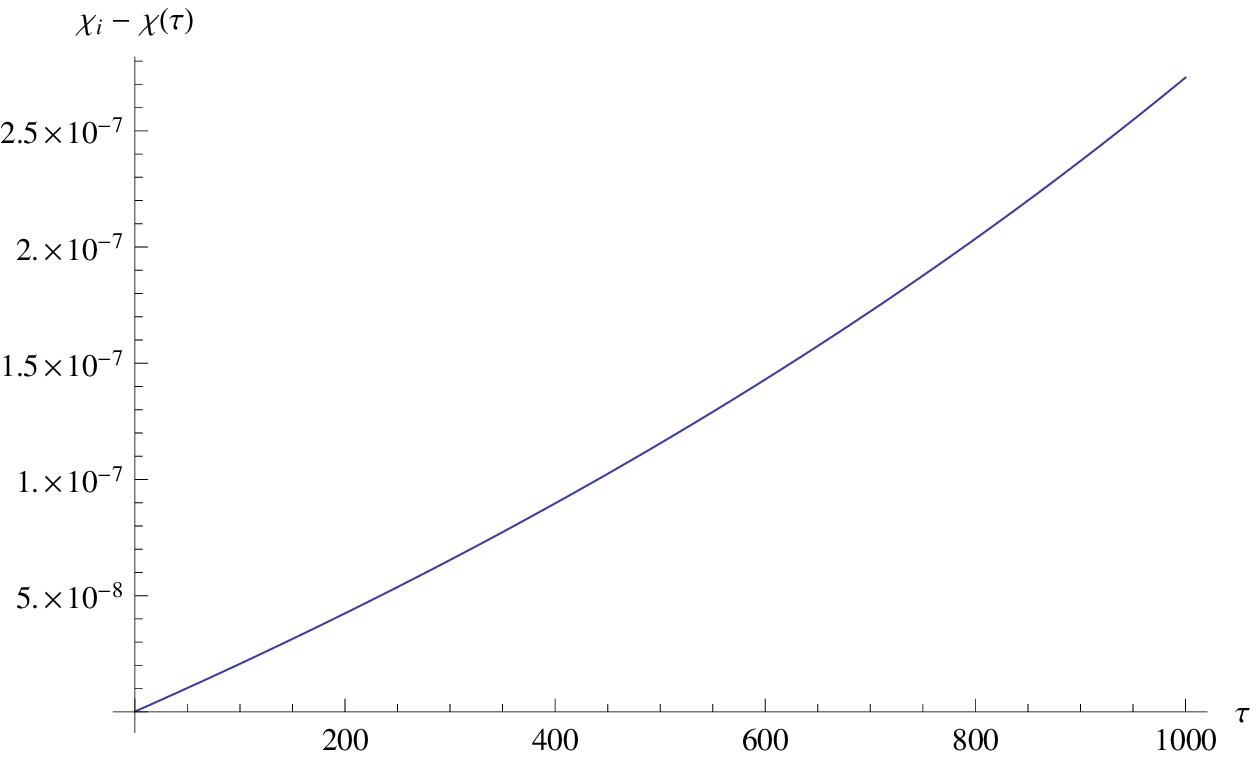}
\hspace{1cm}
\includegraphics[width=6cm,height=4.8cm]{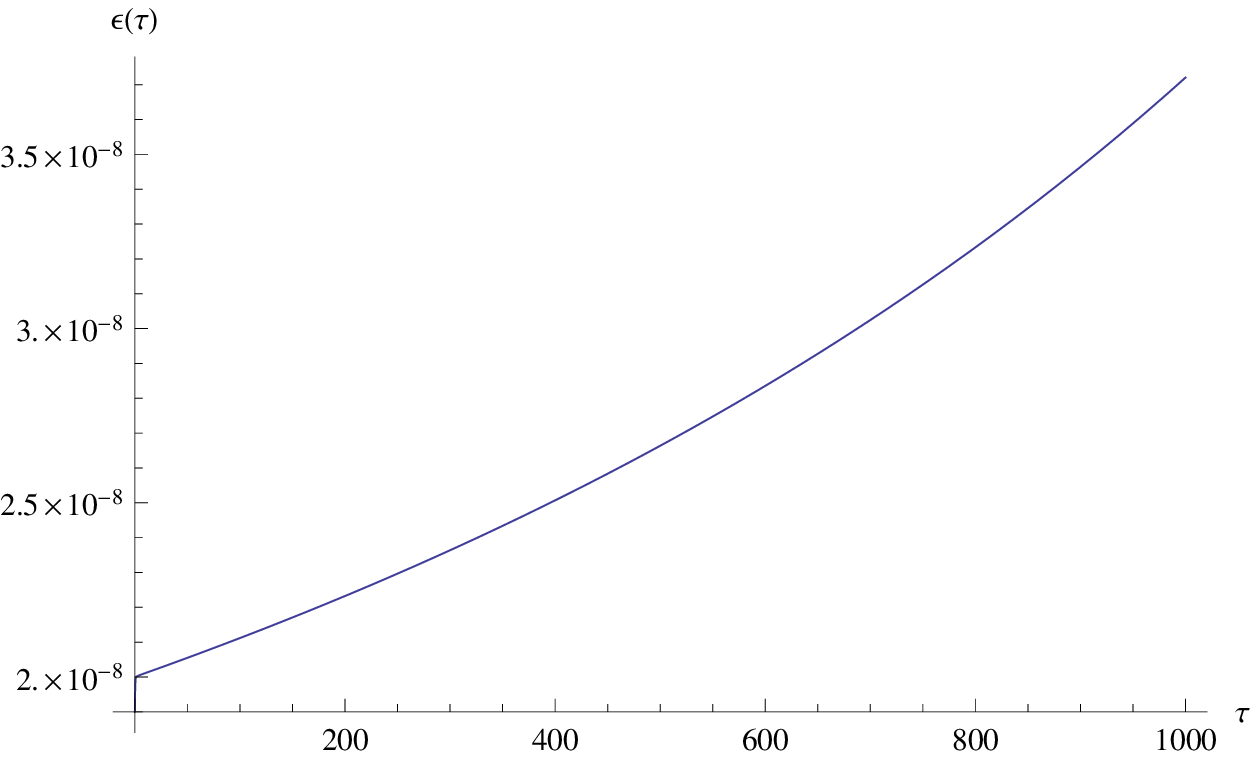}
\caption{\footnotesize Numerical simulation of the geometrical 
quantities $\chi_{\rm in} - \chi(\tau)$ and $\epsilon(\tau)$ 
for $\chi_{\rm in} = \frac1{100}$ and $f(X) = \frac{X}{1-X}$. 
Note the rough agreement with (\ref{geom1000}) during the 
early epoch of nearly de Sitter expansion.}
\label{geom1000}
\end{figure}
 
From relations (\ref{earlyalpha}-\ref{earlydelta}) we 
see that derivatives during the de Sitter epoch take 
the form:
\begin{equation}
\alpha'(\tau) \longrightarrow 
0 \;\; , \;\; 
\beta'(\tau) \longrightarrow 
\frac23 \chi_{\rm in}^2 \;\; , \;\;
\gamma'(\tau) \longrightarrow 
\frac83 \chi_{\rm in}^2 \;\; , \;\;
\delta'(\tau) \longrightarrow 
\frac13 \chi_{\rm in}^2 \; .
\end{equation}
We can also see $\; -\frac12 \beta(\tau) + \delta(\tau) 
\rightarrow \frac12 \chi_{\rm in}^2$. Using these 
relations in the expressions for the Hubble parameter 
and the first slow roll parameter imply:
\begin{equation}
\chi(\tau) \longrightarrow 
\chi_{\rm in} \Bigl( 1 \!-\! 2 \chi_{\rm in}^4 \tau \Bigr)
\qquad , \qquad 
\epsilon(\tau) \longrightarrow 
2 \chi_{\rm in}^4 \; . \label{dSgeom}
\end{equation}
Figure~\ref{geom1000} shows that expressions (\ref{dSgeom}) 
are in rough agreement with numerical simulation.

\begin{figure}[ht]
\includegraphics[width=6cm,height=4.8cm]{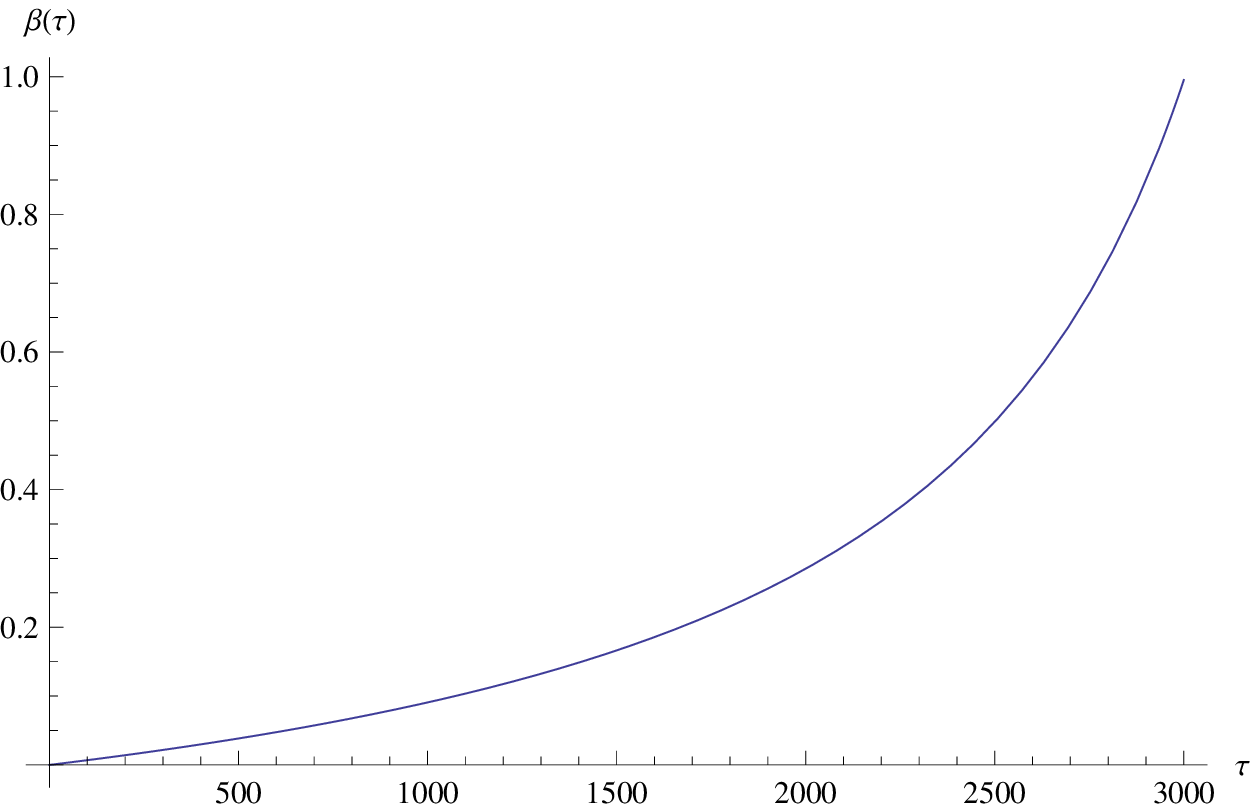}
\hspace{1cm}
\includegraphics[width=6cm,height=4.8cm]{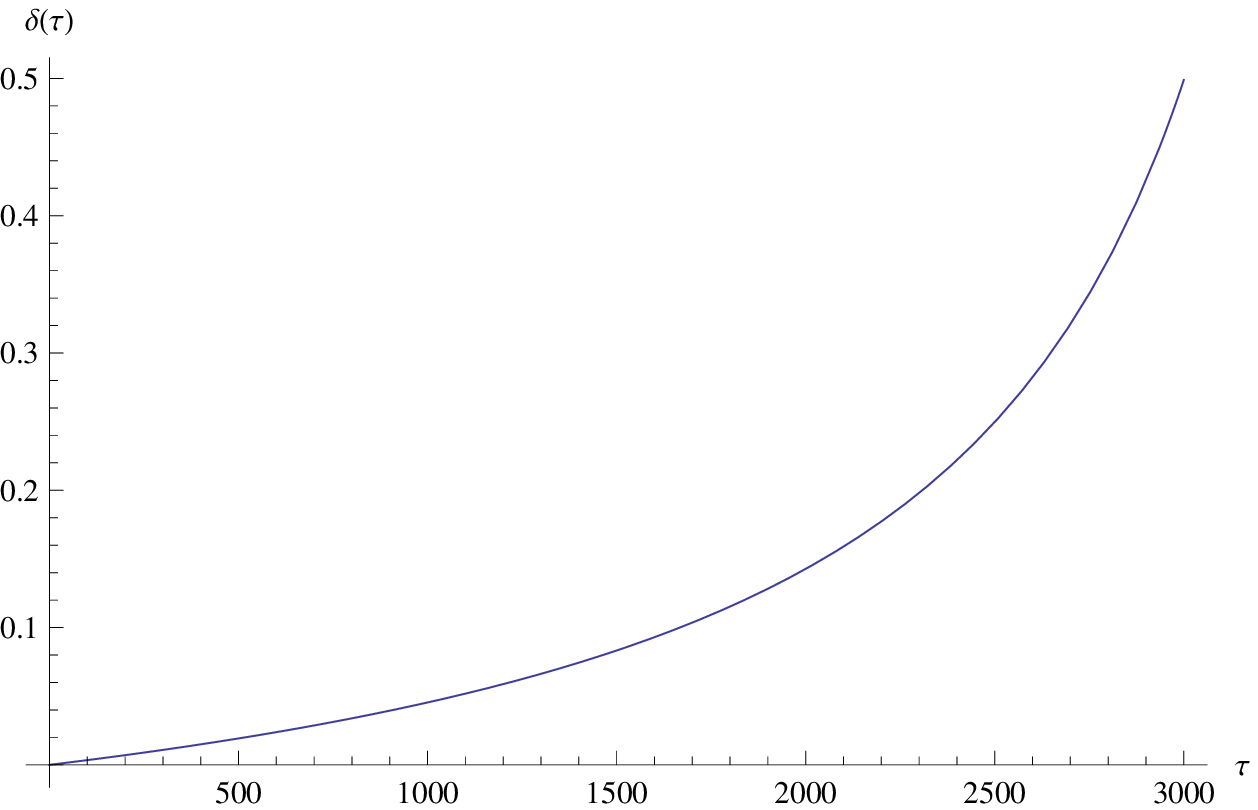}
\caption{\footnotesize Numerical simulation of the auxiliary 
scalar functions $\beta(\tau)$ and $\delta(\tau)$ for 
$\chi_{\rm in} = \frac1{100}$ and $f(X) = \frac{X}{1-X}$. 
Note the curvature quantitative agreement with expressions 
(\ref{earlybeta}) and (\ref{earlydelta}) during the early 
epoch of de Sitter expansion.}
\label{betadelta3000}
\end{figure}

Figure~\ref{alphagamma3500} extends to $\tau = 3500$, 
and shows essentially perfect agreement with expressions 
(\ref{earlyalpha}) and (\ref{earlygamma}). However, 
Figure~\ref{betadelta1000} extends only to $\tau = 1000$, 
and shows a small curvature in addition to the linear 
behaviour predicted by expressions (\ref{earlybeta}) 
and (\ref{earlydelta}). This curvature becomes more 
pronounced for larger values of $\tau$, as is evident 
in Figure~\ref{betadelta3000}. The curvature derives 
from two couplings between the auxiliary scalars which 
are small but not negligible during the de Sitter epoch. 
The first is of $\delta(\tau)$ to $\gamma(\tau)$ (we are 
assuming $f(X) = \frac{X}{1-X}$):
\begin{equation}
\delta'' = 
-3 \frac{\chi}{\chi_{\rm in}} \, \delta' 
+ \chi_{\rm in}^2 \, f(\gamma)
\; \longrightarrow \;
-3 \delta' + \chi_{\rm in}^2 \Bigl(1 \!+\! 2 \gamma\Bigr) 
\; .
\end{equation}
The order $\chi_{\rm in}^4$ correction to $\delta(\tau)$ 
comes from the linear growth of $\gamma(\tau)$ in 
(\ref{earlygamma}):
\begin{equation}
\delta(\tau) \longrightarrow 
\frac13 \chi_{\rm in}^2 \, \tau 
+ \frac89 \chi_{\rm in}^4 \, \tau^2
\; . \label{nextdelta}
\end{equation}
The curvature of $\beta(\tau)$ descends from this 
growth, as reflected in the coupling between $\beta$ 
and $\delta$: 
\begin{equation}
\beta'' = 
-3 \frac{\chi}{\chi_{\rm in}} \, \beta' 
- (2 \!-\! \epsilon) \frac{\chi^2}{\chi_{\rm in}^2} \, \beta 
+ 2 (2 \!-\! \epsilon) \frac{\chi^2}{\chi_{\rm in}^2} \, \delta 
\; \longrightarrow \;
-3 \beta' - 2 \beta + 4 \delta 
\; .
\end{equation}
It follows that the order $\chi_{\rm in}^4$ correction 
to $\beta(\tau)$ is:
\begin{equation}
\beta(\tau) \longrightarrow 
\frac23 \chi_{\rm in}^2 \, \tau 
+ \frac{16}9 \chi_{\rm in}^4 \, \tau^2
\; . \label{nextbeta}
\end{equation}

\begin{figure}[ht]
\includegraphics[width=6cm,height=4.8cm]{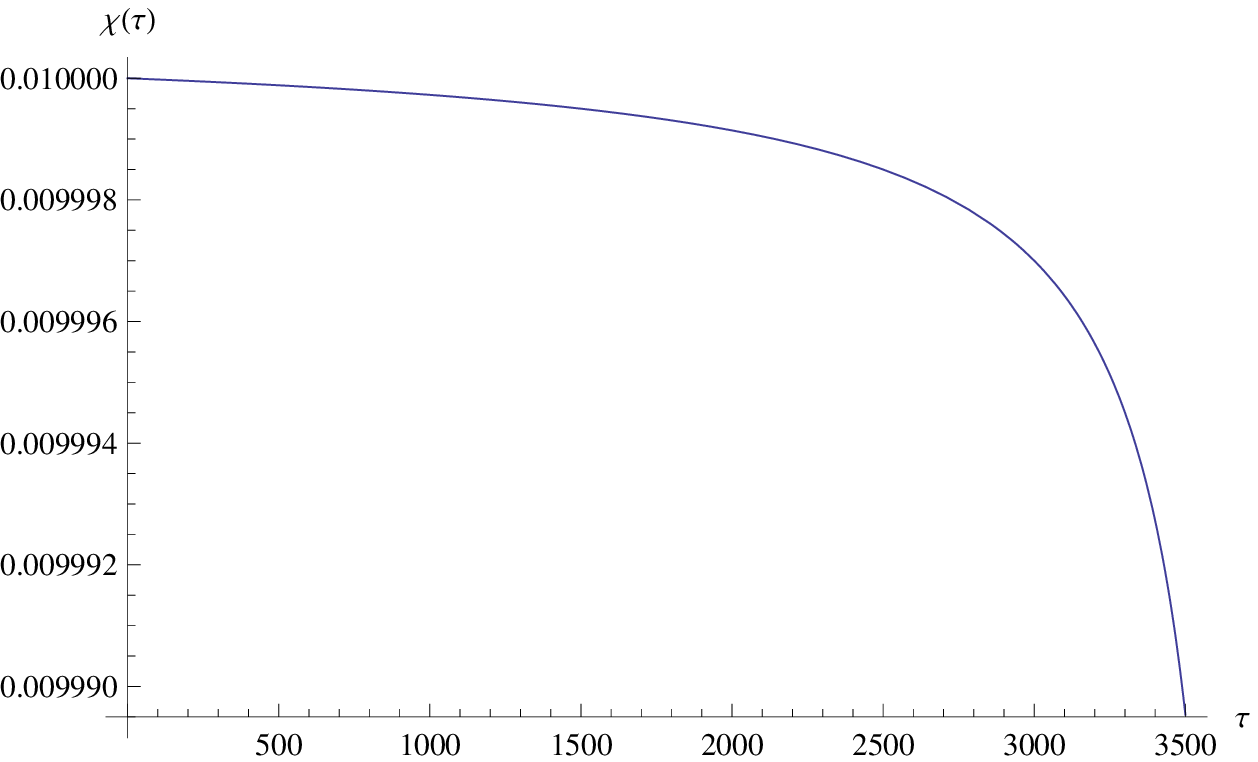}
\hspace{1cm}
\includegraphics[width=6cm,height=4.8cm]{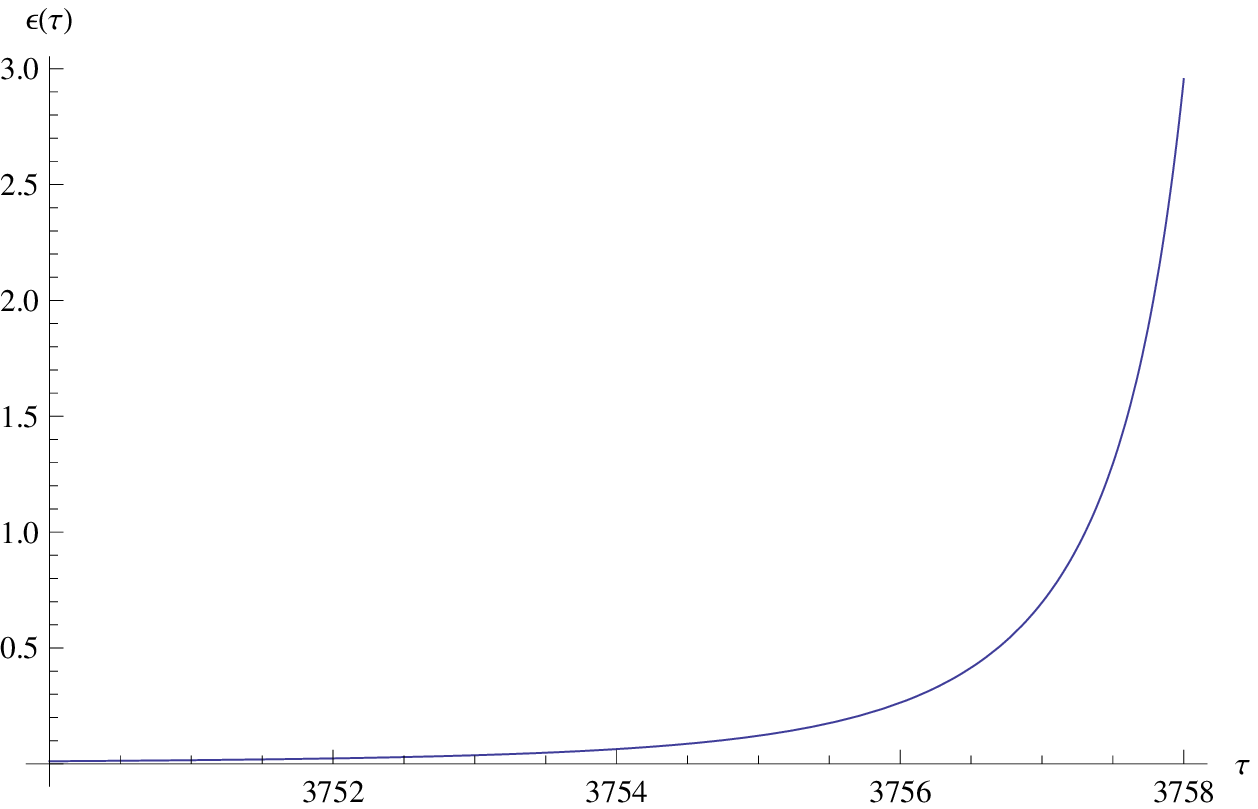}
\caption{\footnotesize Numerical simulation of the Hubble 
parameter and the first slow roll parameter for 
$\chi_{\rm in} = \frac1{100}$ and $f(X) = \frac{X}{1-X}$.}
\label{chi3500}
\end{figure}

\begin{figure}[ht]
\includegraphics[width=6cm,height=4.8cm]{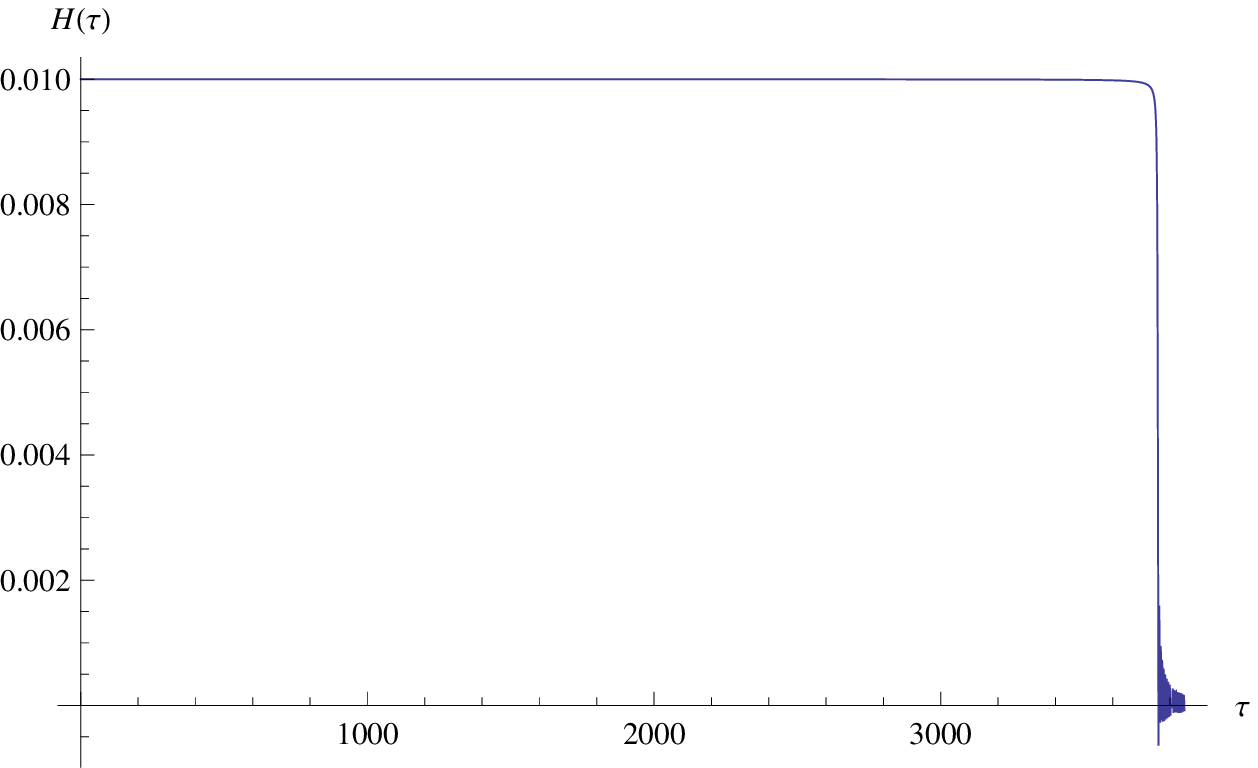}
\hspace{1cm}
\includegraphics[width=6cm,height=4.8cm]{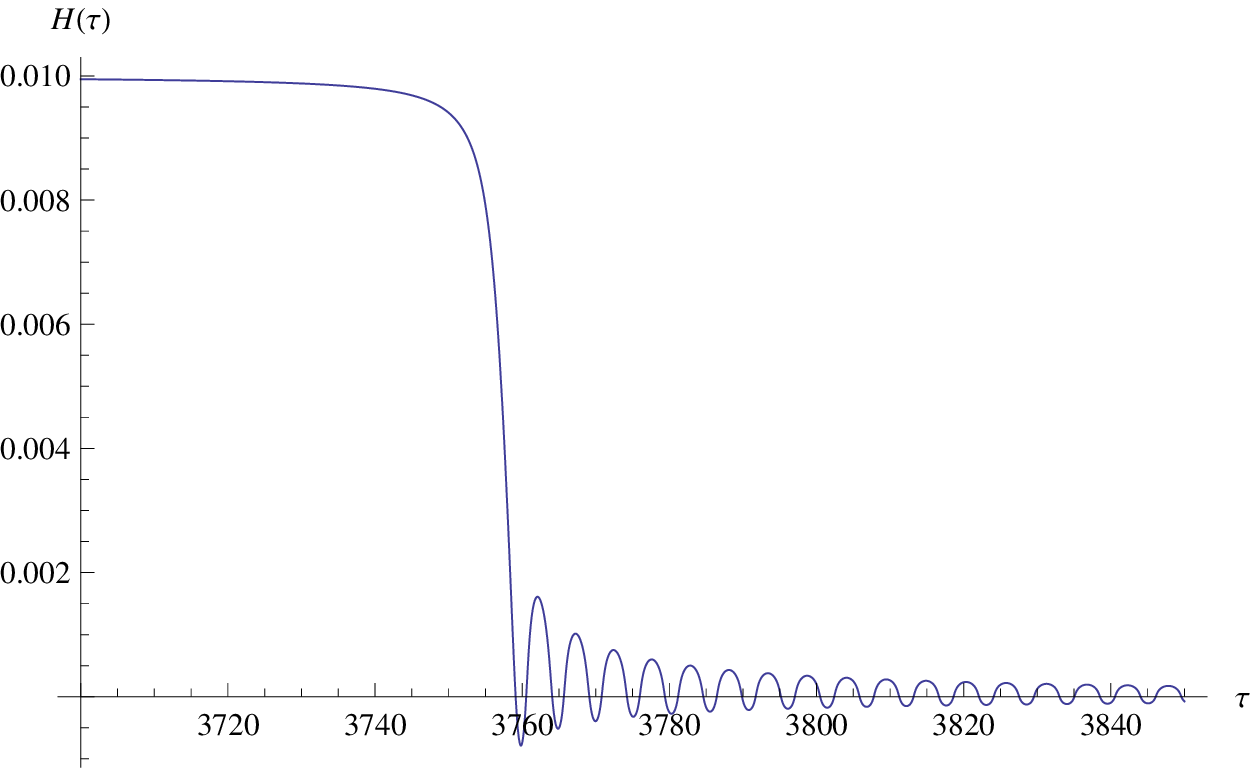}
\caption{\footnotesize Numerical simulation of the Hubble 
parameter for $\chi_{\rm in} = \frac1{100}$ and $f(X) = 
\frac{X}{1-X}$.}
\label{Hubblefull}
\end{figure}

The first effect of the curvature of $\beta(\tau)$ 
and $\delta(\tau)$ is to make $\epsilon(\tau)$ grow 
slightly. That can be seen in the right hand graph of 
Figure~\ref{geom1000}. Curvature also causes the Hubble 
parameter to decline faster than linearly, as can be 
seen in the left hand graph of Figure~\ref{chi3500}. 
From expression (\ref{earlygamma}), and the fact that 
$f(X) = \frac{X}{1-X}$ becomes singular at $X=1$, one 
can estimate that inflation comes to an end at about 
$\tau \simeq \frac38 \chi_{\rm in}^{-2} = 3750$.
The right hand graph of Figure~\ref{chi3500} reveals 
that the actual point where $\epsilon = 1$ is about 
$\tau \simeq 3757.3$. Figure~\ref{Hubblefull} shows 
the Hubble parameter during this period.

\subsection{Reheating and Radiation Domination}

\begin{figure}[ht]
\includegraphics[width=6cm,height=4.8cm]{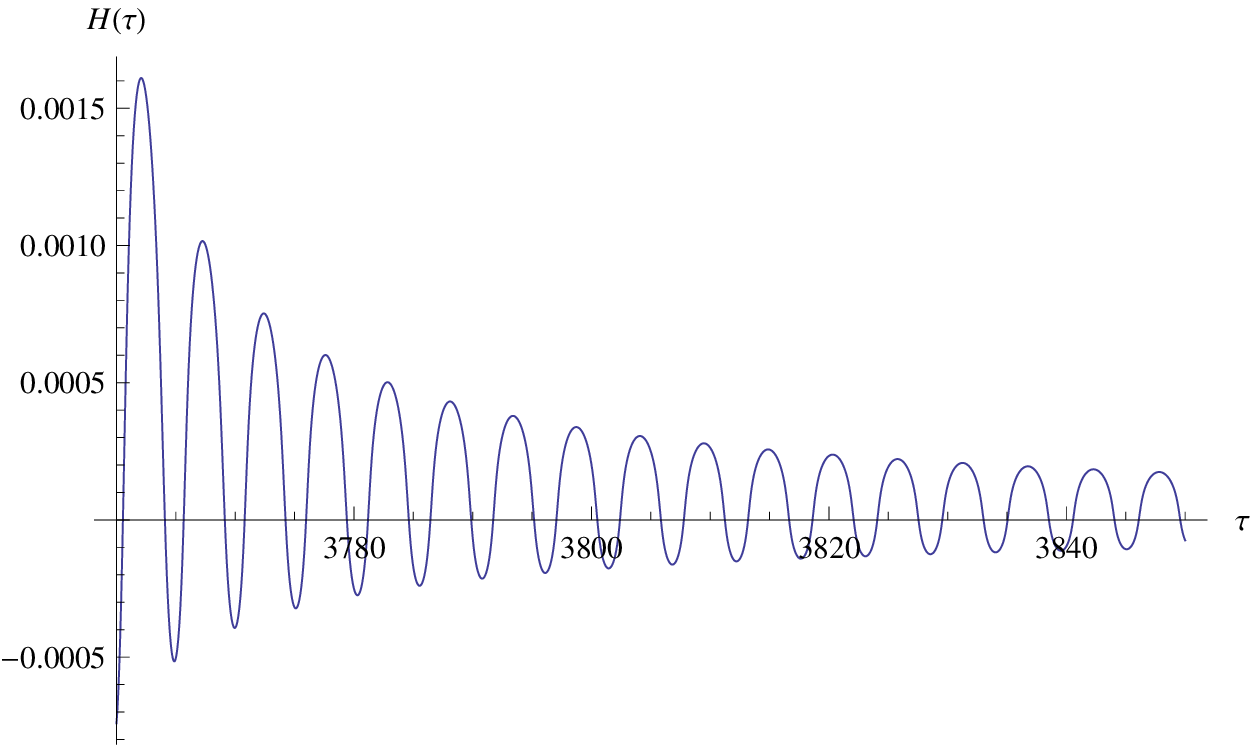}
\hspace{1cm}
\includegraphics[width=6cm,height=4.8cm]{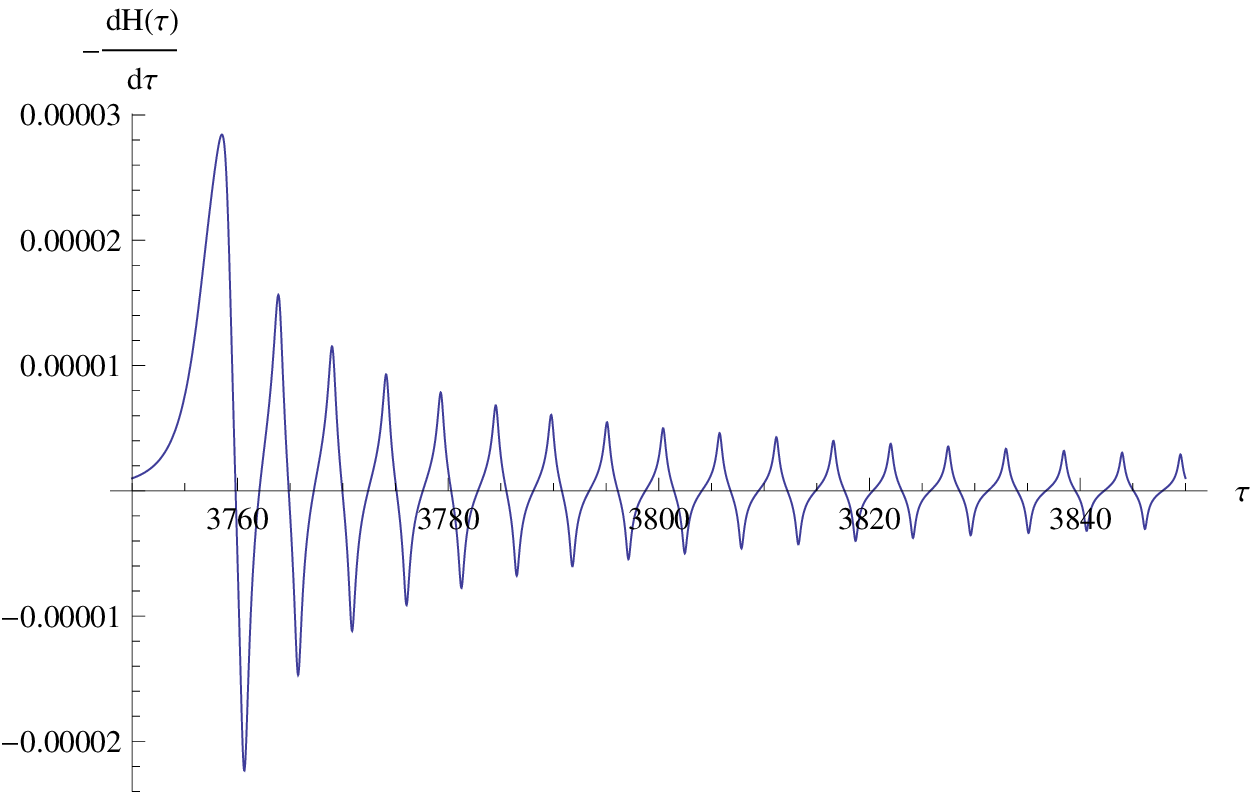}
\caption{\footnotesize Numerical simulation for 
$\chi_{\rm in} = \frac1{100}$ and $f(X) = \frac{X}{1-X}$ 
of the Hubble parameter and its first time derivative 
in the period after the end of inflation.}
\label{Hubbleend}
\end{figure}

Figure~\ref{Hubblefull} shows that screening becomes 
effective quite suddenly and brings inflation to an 
end at about $\tau \simeq 3757.3$. The first slow roll
parameter goes from $\epsilon = 0.3$ to $\epsilon = 1$ 
over a period of only $\Delta \tau \simeq 2$. Thereafter, 
we see from Figure~\ref{Hubbleend} that the Hubble 
parameter oscillates with a decreasing amplitude. 
Of course $\vert \chi(\tau) \vert \gg \chi_{\rm in}$, 
and Hubble friction ceases to be effective. Close 
examination of Figure~\ref{Hubbleend} also reveals 
that the magnitude of $\chi' = -\epsilon \chi^2$ is 
about 50 times larger than $\chi^2$. All of this 
justifies simplifying the auxiliary scalar equations 
accordingly:
\begin{eqnarray}
\alpha'' \, \simeq \, 
+ \frac{\epsilon \chi^2}{\chi_{\rm in}^2} \, \alpha 
\qquad & , & \qquad 
\gamma'' \, \simeq \,
- \frac{2 \epsilon \chi^2}{\chi_{\rm in}^2} \, \alpha 
\; , \label{alphagammasimp1} \\
\delta'' \, \simeq \,
\chi_{\rm in}^2 f'(\gamma) 
\qquad & , & \qquad
\beta'' \, \simeq \,
\frac{\epsilon \chi^2}{\chi_{\rm in}^2} \, 
(\beta \!-\! 2 \delta) 
\; . \label{betadeltasimp1}
\end{eqnarray}

\begin{figure}[ht]
\includegraphics[width=6cm,height=4.8cm]{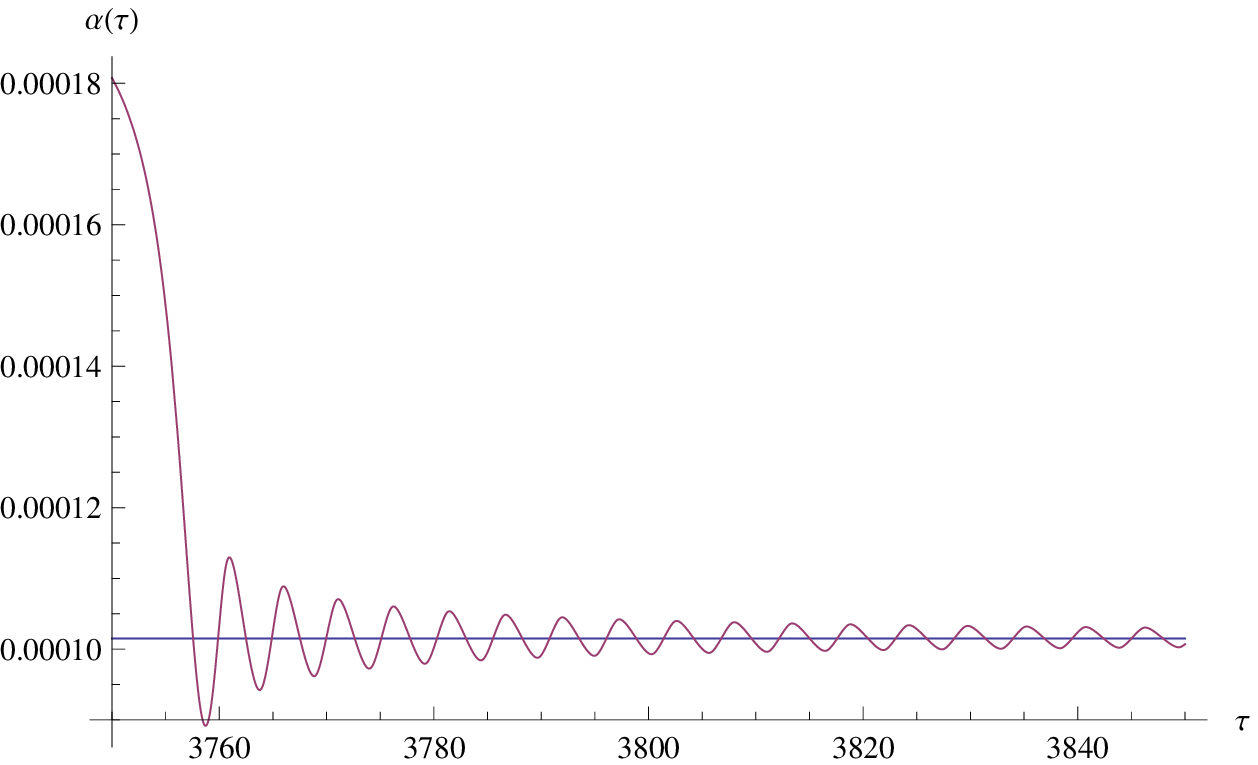}
\hspace{1cm}
\includegraphics[width=6cm,height=4.8cm]{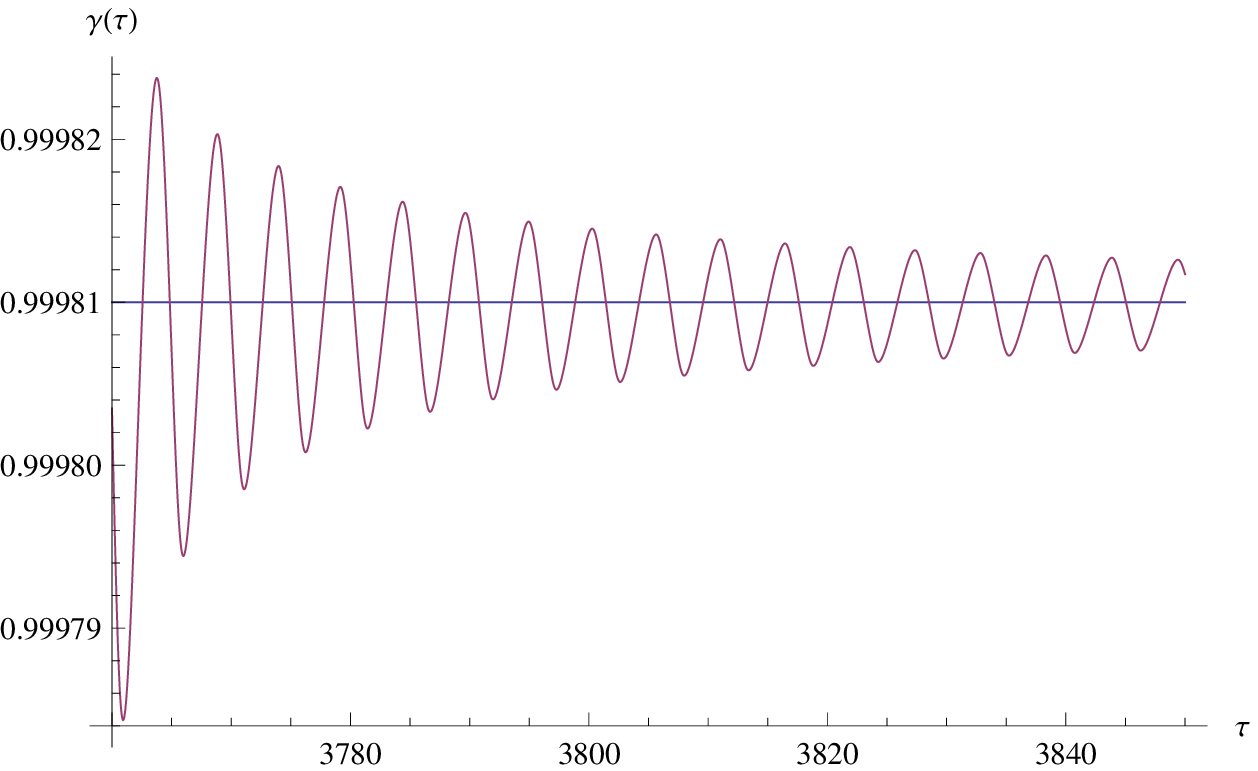}
\caption{\footnotesize Numerical simulation of auxiliary 
scalars $\alpha(\tau)$ and $\gamma(\tau)$ after the end 
of inflation for $\chi_{\rm in} = \frac1{100}$ and $f(X) 
= \frac{X}{1-X}$. The scalar $\alpha(\tau)$ (on the left) 
oscillates around $\alpha_0 \simeq 0.0001015$. The scalar 
$\gamma(\tau)$ (on the right) oscillates around $\gamma_0 
\simeq 0.99981$.}
\label{Alphagammaend}
\end{figure}

Figure~\ref{Alphagammaend} shows that $\alpha(\tau)$ 
and $\gamma(\tau)$ experience oscillations of decreasing 
amplitude about central values of $\alpha_0 \simeq 0.0001015$ 
and $\gamma_0 \simeq 0.99981$, respectively. This means 
one can carry the simplifications of the equations 
(\ref{alphagammasimp1}) a step further:
\begin{eqnarray}
& \mbox{} &
\Big( \, \alpha'' \simeq
\alpha_0 \frac{\epsilon \chi^2}{\chi_{\rm in}^2} 
\quad \& \quad
\gamma'' \simeq
-2\alpha_0 \frac{\epsilon \chi^2}{\chi_{\rm in}^2} 
\, \Big)
\;\; \Longrightarrow 
\nonumber \\
& \mbox{} & 
\hspace{1cm} 
\Delta \alpha(\tau) \, \equiv \,
\alpha(\tau) \!-\! \alpha_0 
\, \simeq \,
-\frac12 \Bigl[ \gamma(\tau) \!-\! \gamma_0 \Bigr]
\, \equiv \,
-\frac12 \Delta \gamma(\tau) 
\; . \label{alphagammasimp2}
\end{eqnarray}
At this stage we can gain a rough understanding of 
what is driving the oscillations. Recall expression 
(\ref{chieps}) for $\epsilon \chi^2$ and use $\chi^2 
\ll \chi_{\rm in}^2$ as well as $2 \alpha' + \gamma' 
\simeq 0$ to simplify the numerator of (\ref{chieps}):
\begin{eqnarray}
\lefteqn{
\Bigl[\alpha f'(\gamma) \!+\! 2 f(\gamma)\Bigr] \chi_{\rm in}^4
\!-\! \frac43 \chi_{\rm in}^2 
\!+\! (2 \alpha' \!+\! \gamma') \delta' \chi_{\rm in}^2
\!+\! \frac43 \chi^2 
\!+\! 6 (-\beta \!+\! 2 \delta) \chi^4 } 
\nonumber \\
& & \hspace{1cm} 
\simeq \Bigl[ \alpha f'(\gamma) \!+\! 2 f(\gamma) \Bigr] 
\chi_{\rm in}^4 
\!-\! \frac43 \chi_{\rm in}^2 
\; , \\
& & \hspace{1cm} 
\simeq \Bigl[ \alpha_0 f'(\gamma_0) \!+\! 2 f(\gamma_0) \Bigr] 
\chi_{\rm in}^4 
\!-\! \frac43 \chi_{\rm in}^2 
+ \Bigl[ \frac32 f'(\gamma_0) 
\!+\! \alpha_0 f''(\gamma_0) \Bigr] \chi_{\rm in}^4 
\!\times\! \Delta \gamma 
\; , \qquad \\
& & \hspace{1cm} 
\simeq 0.711 \!\times\! \Delta \gamma 
\; . \label{finalsimp}
\end{eqnarray}
Substituting (\ref{finalsimp}) into (\ref{chieps}), 
and then into (\ref{alphagammasimp2}) gives what 
is a recognizable oscillator equation for 
$\Delta \gamma(\tau)$:
\begin{equation}
\Delta \gamma'' \, \simeq \,
- \frac{4.333 \!\times\! \Delta \gamma}
{2 \!+\! 12 \beta' \, \chi_{\rm in} \, \chi 
\!+\! 12 (-\beta \!+\! 2 \delta) \chi^2} 
\; . \label{osceqn}
\end{equation}
Of course this also implies oscillations for 
$\Delta \alpha \simeq -\frac12 \Delta \gamma$, 
and for $\epsilon \chi^2$. The decreasing amplitude 
of oscillation is presumably due to the residual 
effect of Hubble friction.

\begin{figure}[ht]
\includegraphics[width=6cm,height=4.8cm]{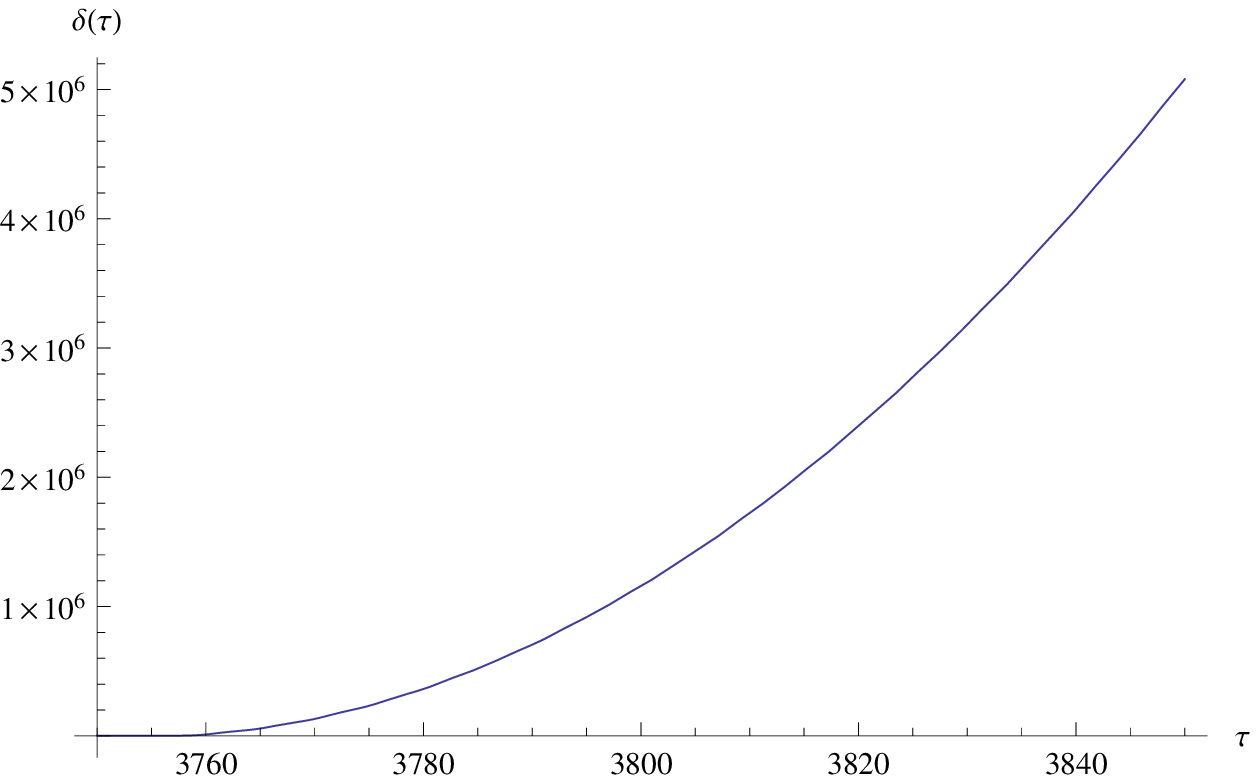}
\hspace{1cm}
\includegraphics[width=6cm,height=4.8cm]{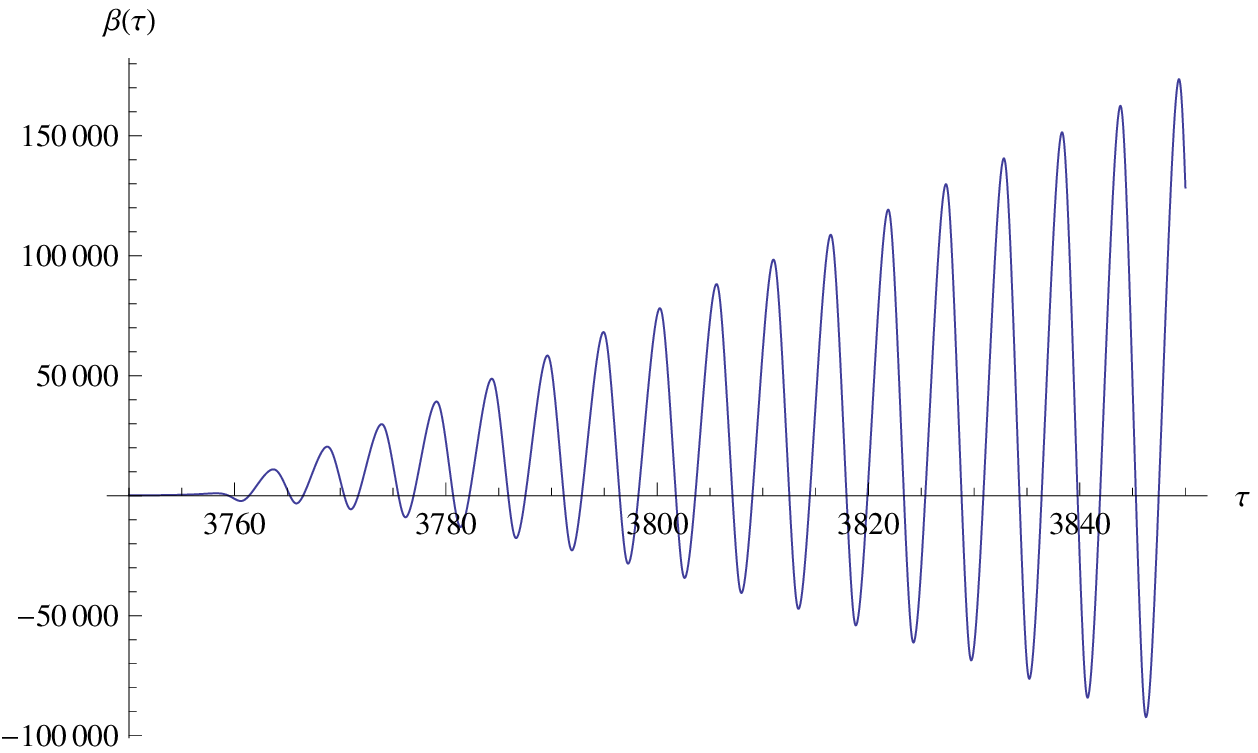}
\caption{\footnotesize Numerical simulation of auxiliary 
scalars $\delta(\tau)$ and $\beta(\tau)$ after the end of 
inflation for $\chi_{\rm in} = \frac1{100}$ and $f(X) = 
\frac{X}{1-X}$. The scalar $\delta(\tau)$ (on the left) 
grows, roughly like $(\tau - 2760)^{1.82}$. The scalar 
$\beta(\tau)$ (on the right) oscillates with a linearly 
increasing central value and a linearly increasing amplitude.}
\label{Deltabetaend}
\end{figure}

Figure~\ref{Deltabetaend} shows the auxiliary scalars 
$\delta(\tau)$ and $\beta(\tau)$. We can understand the 
growth of $\delta(\tau)$ by making a further simplification 
of its equation (\ref{betadeltasimp1}):
\begin{equation}
\delta'' \, \simeq \,
\chi_{\rm in}^2 \, f'(\gamma_0) \, \simeq \, 
2770 
\; . \label{deltasimp2}
\end{equation}
That would give quadratic growth. What Figure~\ref{Deltabetaend} 
actually shows is somewhat slower growth, $\delta(\tau) \simeq 
\frac{2770}2 \times (\tau - 2760)^{1.82}$. The reduction 
seems to be due to the residual effect of Hubble friction. 
We can understand the behaviour of $\beta(\tau)$ by making 
a similar simplification of its equation (\ref{betadeltasimp1}):
\begin{equation}
\beta'' \simeq 
- \frac{2 \epsilon \chi^2}{\chi_{\rm in}^2} \, \delta 
\; . \label{betasimp2}
\end{equation}
The source on the right hand side oscillates and grows 
linearly, so the response of $\beta(\tau)$ in 
Figure~\ref{Deltabetaend} can be understood by stripping 
away all the constants:
\begin{equation}
\left\{{f''(x) = -x \cos(x) \atop f(0) = 0 = f'(0)} \right\}
\qquad \Longrightarrow \qquad
f(x) = x \Bigl[ \cos(x) \!+\! 1\Bigr] - 2 \sin(x) 
\; . \end{equation}

The preceding analysis and numerical results have dealt 
with the period immediately after the end of inflation. 
A point of great significance is that {\it the Hubble 
parameter becomes negative}. This is evident in 
Figure~\ref{Hubbleend}. Of course negative $H$ means 
that the universe is contracting, which must concentrate 
whatever matter particles are produced by the fluctuating 
geometry. There will be a similar concentration of the 
last graviton and inflaton perturbations to have been 
generated during inflation. We are not now in a position 
to analyze this process in detail but it seems obvious 
that rapid reheating will occur. And note that the onset 
of radiation domination, with $\epsilon = 2$, turns off 
the source for further evolution of the key auxiliary 
scalar $\gamma(\tau)$. Assuming that this point is 
reached, the screening effect goes quiescent with $\gamma
= \gamma_*$, and the universe experiences the usual phase 
of radiation domination with essentially zero cosmological 
constant.

\subsection{Late Time Acceleration}

We turn now to the time $t= t_m$, long after reheating, 
when the universe makes the transition to matter domination. 
At that point $\gamma$ begins to evolve again. This will 
induce a corresponding change in our non-local source
(\ref{DL}):
\begin{eqnarray}
\Lambda^2 \, h(GC) &\!\! = \!\!&
\Lambda^2 f(\gamma) \, = \,
\Lambda^2 \, f\Bigg(
G \, \frac19 \, \frac{1}{\square} \, R \, \frac{1}{\square_c}
\Big[ \frac13 \, R^2 - R_{\mu\nu} R^{\mu\nu} \Big]
\Bigg)
\nonumber \\
&\!\! \simeq \!\!&
\Lambda^2 f(\gamma_*) 
+ \Lambda^2 f'(\gamma_*) \!\times\! 
\Bigl[ \gamma \!-\! \gamma_* \Bigr] 
\; . \label{Lambdachange}
\end{eqnarray}
Let us first examine the change in $[\gamma - \gamma_*](t)$,
returning to co-moving time as an evolution variable:
\begin{eqnarray}
\lefteqn{ \Bigl[ \gamma \!-\! \gamma_*\Bigr](t) 
\, = \,
\frac{G}{9} \int_{t_m}^{t} \!\! \frac{dt'}{a^3(t')} \! 
\int_{t_m}^{t'} \!\! dt'' a^3(t'') \!\times\! 
6 [2 \!-\! \epsilon(t'')] H^2(t'') } 
\nonumber \\
& & \hspace{3cm} 
\times\! \frac1{a(t'')} \!\int_{t_m}^{t''} \!\! 
\frac{dt'''}{a(t''')} \! 
\int_{t_m}^{t'''} \!\! dt'''' a^2(t'''') \!\times\! 
12 [1 \!-\! \epsilon] H^4(t'''') 
\; . \qquad \label{Dgammamatter}
\end{eqnarray}
To simplify the discussion, we assume that the universe 
is perfectly matter dominated after $t = t_m$:
\begin{equation}
t > t_m 
\qquad \Longrightarrow \qquad 
a(t) = a_m \Bigl( \frac{t}{t_m} \Bigr)^{\frac23} 
\quad , \quad 
H(t) = \frac{2}{3 t} 
\quad , \quad
\epsilon(t) = \frac32 
\; . \label{matterdom}
\end{equation}
Substituting (\ref{matterdom}) into (\ref{Dgammamatter}) 
and performing the trivial integrations gives:
\begin{equation}
\Bigl[ \gamma \!-\! \gamma_*\Bigr](t) 
\, = \,
-\frac{2^7 G}{3^3 \, 5 \, t_m^2} \Biggl\{ 
\frac56 \!-\! 3 \Bigl( \frac{t}{t_m}\Bigr)^{\frac13} 
\!+\! \frac{15}{4} \Bigl( \frac{t_m}{t} \Bigr)^{\frac23} 
\!-\! \frac53 \frac{t_m}{t} 
\!+\! \frac1{12} \Bigl( \frac{t_m}{t} \Bigr)^2 \Biggr\} 
\; .
\end{equation}
It is more useful to express this in terms of the 
Hubble parameter at the time of matter domination 
$H_m \equiv H(t_m)$, and to take the late time limit:
\begin{equation}
\lim_{t \gg t_m} \Bigl[ \gamma \!-\! \gamma_* \Bigr](t) 
\, = \,
-\frac{16}{9} \, G H_m^2 
\; .
\end{equation}

Let us now turn to the question of what sort of function 
$f(X)$ would give a late time cosmological constant of 
the right size. The two terms on the right hand side of 
(\ref{Lambdachange}) have different roles: \\
${\bullet \;}$ The first cancels the large, bare 
cosmological constant:
\begin{equation}
\frac{6 H_{\rm in}^2}{16 \pi G} = 
\Lambda^2 f(\gamma_*) 
\; ; \label{cond1}
\end{equation}
${\bullet \;}$ The second supplies the small, positive 
cosmological constant needed to cause the observed late 
time acceleration:
\begin{equation}
\frac{6 H_0^2}{16 \pi G} 
\, = \,
\Lambda^2 f'(\gamma_*) \!\times\! 
\frac{16}{9} \, G H_m^2 
\; . \label{cond2}
\end{equation}
Taking the ratio of (\ref{cond2}) to (\ref{cond1}) 
implies $f(X)$ must obey:
\begin{equation}
\frac{f'(\gamma_*)}{f(\gamma_*)} 
= 
\frac{9}{16} \Bigl( \frac{H_0}{H_{\rm in}} \Bigr)^2
\frac{1}{G H_m^2} 
\; . \label{ratio}
\end{equation}
Some of the numbers in expression (\ref{ratio}) 
are known:
\begin{equation}
G H_m^2 
\, \simeq \,
10^{10} \!\times G H_0^2 
\, \simeq \, 
10^{-112} 
\; .
\end{equation}
If we assume $H_{\rm in} \simeq 10^{55} H_0$, 
the result is:
\begin{equation}
\frac{f'(\gamma_*)}{f(\gamma_*)} 
\simeq 10^{2} 
\; . \label{estimate}
\end{equation}

Conditions (\ref{cond1}) and (\ref{estimate}) are 
certainly not obeyed for the simple ansatz $f(X) = 
\frac{X}{1 - X}$ that was used for our numerical 
simulations. We therefore consider a 1-parameter 
family of more singular models:
\begin{equation}
f(X) = \frac1{(1 \!-\! X)^{\omega}} - 1 
\qquad \Longrightarrow \qquad
f'(X) = \frac{\omega}{(1 \!-\! X)^{\omega + 1}} 
\; .
\end{equation}
Because $\gamma_*$ is very close to unity we can express 
conditions (\ref{cond1}) and (\ref{estimate}) as:
\begin{equation}
\frac1{(1 \!-\! \gamma_*)^{\omega}} 
\simeq 10^{10} 
\qquad , \qquad
\frac{\omega}{1 \!-\! \gamma_*} 
\simeq 10^{2} 
\; . \label{omegacond}
\end{equation}
An approximate solution is clearly $\omega \simeq 5$. 
Different assumptions about $H_i$ can be accommodated 
with only small changes in the exponent $\omega$. Note 
also that because the actual expansion history is not 
(\ref{matterdom}), the current phase of acceleration 
will eventually end, although after a very long time.

\section{Epilogue}

The model presented in this paper was ultimately 
motivated by the fact that gravitation is the
dominant force responsible for the evolution of
the universe. It is therefore reasonable to seek
a model exclusively using gravitational degrees
of freedom. Its construction was dictated by
consistency with perturbative results as well 
as with satisfaction of basic cosmological
requirements. Alternatively, its construction
can be viewed as simply an {\it ansatz} whose
implications should be studied. 

These implications include an end by gravitational
means of an inflationary era of adequate duration, 
an oscillatory era that follows and can lead a
naturally reheated universe to the epoch of radiation 
domination and, thereafter, to matter domination. 
The analysis is based on numerical and semi-analytical 
methods.

Finally, the several conjectures we have had to 
make in this analysis should not be allowed to 
obscure the fact that this model provides natural 
explanations both for why the current phase of 
acceleration happens so late in cosmological
history, and for why its source appears to be 
an absurdly small, positive cosmological constant. 
Our key point is that late time acceleration is not 
the result of a small bare cosmological constant 
but rather of a very small fractional change in 
quantum gravitational screening which was triggered 
by the transition from radiation domination to 
matter domination.

\newpage

\centerline{\bf Acknowledgements}

This work was partially supported by the European Union's 
Seventh Framework Programme (FP7-REGPOT-2012-2013-1) under 
grant agreement number 316165; by the European Union's Horizon 
2020 Programme under grant agreement 669288-SM-GRAV-ERC-2014-ADG;
by NSF grant PHY-1506513; and by the Institute for Fundamental 
Theory at the University of Florida.


\vspace{1cm}

\begin{thebibliography}{99}

\bibitem{Planck2015} 
``Planck 2015 results. XIII. Cosmological parameters''
(June 2016, v3),
{\bf arXiv:}1502.01589 [astro-ph]

\bibitem{riess}
``A 2.4\% Determination of the Local Value of the Hubble 
Constant'' (June 2016, v3),
{\bf arXiv:}1604.01424 [astro-ph]

\bibitem{Linde} A. Linde, 
{\it Particle Physics and Inflationary Cosmology}
(Harwood, Amsterdam, 1990).

\bibitem{Slava} V. F. Mukhanov, 
{\it Physical Foundations of Cosmology}
(Cambridge University Press, Cambridge, 2005).

\bibitem{gravitons}
L. P. Grishchuck, 
Sov. Phys. JETP {\bf 40} (1975) 409; \\
L. H. Ford and L. Parker, 
Phys. Rev. {\bf D16} (1977) 1601.

\bibitem{NctRpw1} N. C. Tsamis and R. P. Woodard, \\
Nucl. Phys. {\bf B474} (1996) 235, 
{\bf arXiv:}hep-ph/9602315; \\
Annals Phys. {\bf 253} (1997) 1, 
{\bf arXiv:}hep-ph/9602316; \\
Int. J. Mod. Phys. {\bf 20} (2011) 2847,
{\bf arXiv:}1103.5134 [gr-qc]

\bibitem{NctRpw2} N. C. Tsamis and R. P. Woodard, 
Nucl. Phys. {\bf B724} (2005) 295, \\
{\bf arXiv:}gr-qc/0505115. \\
T. Prokopec, N. C. Tsamis and R. P. Woodard,
Annals Phys. {\bf 323} (2008) 1324, 
{\bf arXiv:}0707.0847 [gr-qc]


\newpage

\bibitem{nonloc} 
L. Parker and D. J. Toms, Phys. Rev. {\bf D32} (1985) 1409;
T. Banks, Nucl. Phys. {\bf B309} (1988) 493; C. Wetterich, Gen. Rel.
Grav. {\bf 30} (1998) 159, gr-qc/9704052;
A. O. Barvinsky, Phys. Lett. {\bf B572} (2003) 109, hep-th/0304229;
D. Espriu, T. Multamaki and E. C. Vagenas, Phys. Lett. {\bf B628}
(2005) 197, gr-qc/0503033; H. W. Hamber and R. M. Williams, Phys.
Rev. {\bf D72} (2005), 044026, hep-th/0507017; T. Biswas, A.
Mazumdar and W. Siegel, JCAP {\bf 0603} (2006) 009, hep-th/0508194;
D. Lopez Nacir and F. D. Mazzitelli, Phys. Rev. {\bf D75} (2007)
024003, hep-th/0610031; J. Khoury, Phys. Rev. {\bf D76} (2007)
123513, hep-th/0612052; S. Capozziello, E. Elizalde, S. Nojiri
and S. D. Odintsov, Phys. Lett. {\bf B671} (2009) 193,
arXiv:0809.1535; F. W. Hehl and B. Mashhoon, Phys. Lett. {\bf B673} (2009)
279, arXiv:0812.1059; Phys. Rev. {\bf D79} (2009) 064028, arXiv:0902.0560;
T. Biswas, T. Koivisto and A. Mazumdar, JCAP {\bf 1011} 
(2010) 008, arXiv:1005.0590; S. Nojiri, S. D. Odintsov, M. Sasaki 
and Y.-l. Zhang, Phys. Lett. {\bf B696} (2011) 278,arXiv:1010.5375;
K. Bamba, S. Nojiri, S. D. Odintsov and M. Sasaki, Gen. Rel. Grav. 
{\bf 44} (2012) 1321, arXiv:1104.2692; Y.-l. Zhang and M. Sasaki, Int. 
J. Mod. Phys. {\bf D21} (2012) 1250006, arXiv:1108.2112; A. O. Barvinsky, 
Phys. Lett. {\bf b710} (2012) 12, arXiv:1107.1463; Phys. Rev. {\bf D85} 
(2012) 104018, arXiv:1112.4340; E. Elizalde, E. O. Pozdeeva and S. Y.
.Vernov, Phys. Rev. {\bf D85} (2012) 044002, arXiv:1110.5806; A. O.
Barvinsky and Y. V. Gusev, Phys. Part. Nucl. {\bf 44} (2013) 213,
arXiv:1209.3062; T. Biswas, A. Conroy, A. S. Koshelev and A. Mazumdar,
Class. Quant. Grav. {\bf 31} (2013) 015022 (2013), arXiv:1308.2319;
P. G. Ferreira and A. L. Maroto, Phys. Rev. {\bf D88} (2013) 123502,
arXiv:1310.1238; S. Foffa, M. Maggiore and E. Mitsou, arXiv:1311.3421; 
arXiv:1311.3435; E. O. Pozdeeva and S. Y. Vernov, arXiv:1401.7550; A. 
Kehagias and M. Maggiore, JHEP {\bf 08} (2014)029, arXiv:1401.8289; 
M. Maggiore and M. Mancarella, arXiv:1402.0448; J. F. Donoghue and 
B. K. El-Menoufi, arXiv:1402.3252; Y. Dirian, S. Foffa, N. Khosravi, 
M. Kunz and M. Maggiore, arXiv:1403.6068; A. Conroy, T. Koivisto, 
A. Mazumdar and A. Teimouri, Class. Quant. Grav. {\bf 32} (2015) 015024,
arXiv:1406.4998; Y. Dirian and E. Mitsou, JCAP {\bf 1410}, (2014) 065, 
arXiv:1408.5058; B. Mashhoon, Galaxies {\bf 3} (2015) 1, arXiv:1411.5411;
E. Mitsou, arXiv:1504.04050; J. F. Donoghue and B. K. El-Menoufi, JHEP 
{\bf 1510} (2015) 044, arXiv:1507.06321; B. K. El-Menoufi, JHEP {\bf 1605}, 
(2016) 035, arXiv:1511.08816; T. Bautista and A. Dabholkar, arXiv:1511.07450, 
T. Bautista, A. Benevides, A. Dabholkar and A. Goel, arXiv:1512.03275;
Y. l. Zhang, K. Koyama, M. Sasaki and G. B. Zhao, JHEP {\bf 1603} (2016)
039, arXiv:1601.03808; H. Nersisyan, Y. Akrami, L. Amendola, 
T. S. Koivisto and J. Rubio, arXiv:1606.04349.

\bibitem{NctRpw3} N. C. Tsamis and R. P. Woodard, 
Phys. Rev. {\bf D80} (2009) 083512, \\
{\bf arXiv:}0904.2368 [gr-qc]

\bibitem{NctRpw4} N. C. Tsamis and R. P. Woodard, 
Phys. Rev. {\bf D81} (2010) 103509, \\
{\bf arXiv:}1001.4929 [gr-qc] \\
M. G. Romania, N. C. Tsamis and R. P. Woodard, 
Lect. Notes Phys. {\bf 863} (2013) 375, 
{\bf arXiv:}1204.6558 [gr-qc] 

\bibitem{riess2} A. G. Riess {\it et al.}, 
Astron. J. {\bf 116} (1998) 1009, \\
{\bf arXiv:}astro-ph/9805201. \\
S. Perlmutter {\it et al.},
Astrophys. J. {\bf 517} (1999) 565, \\
{\bf arXiv:}astro-ph/9812133.

\bibitem{wang} Y. Wang and P. Mukherjee, 
Astrophys. J. {\bf 650} (2006) 1, \\
{\bf arXiv:}astro-ph/0604051. \\
U. Alam, V. Sahni and A. A. Starobinsky, 
JCAP {\bf 0702} (2007) 011, \\
{\bf arXiv:}astro-ph/0612381.

\bibitem{deser} S. Deser and R. P. Woodard,
JCAP {\bf 1311} (2013) 036, \\
{\bf arXiv:}1307.6639 [astro-ph] \\
R. P. Woodard,
Found. Phys. {\bf 44} (2014) 213, 
{\bf arXiv:}1401.0254 [astro-ph]

\bibitem{odintsov} S. Nojiri and S. D. Odintsov,
Phys. Lett. {\bf B659} (2008) 821, \\
{\bf arXiv:}0708.0924 [hep-th] 

\bibitem{NctRpw5} N. C. Tsamis and R. P. Woodard, 
JCAP {\bf 1409} (2014) 008, \\
{\bf arXiv:}1001.4929 [astro-ph] 


\end{thebibliography}
\end{document}